\documentclass[twocolumn,aps,superscriptaddress,nofootinbib,floatfix,linenumbers]{revtex4}
\usepackage{bm}
\usepackage{bbm}
\usepackage{graphicx} % Required for inserting images
\usepackage[utf8]{inputenc}
\usepackage[dvipsnames]{xcolor}
\usepackage{amsmath}
\usepackage{amssymb}
\usepackage{lipsum}
\usepackage{ulem}
\usepackage{float}
\usepackage{amsfonts}
\usepackage{physics}
\usepackage{slashed}
\usepackage{wasysym}
\usepackage[hidelinks]{hyperref}
\usepackage{makecell}
\usepackage[thinlines]{easytable}
\usepackage[caption=false]{subfig}
\usepackage{tabularray}
\NewColumnType{M}[1]{Q[m,c,#1]} 
\usepackage{breakcites}
\usepackage{feynmp-auto} % instead of feynmf
\usepackage{comment}
\begin{document}

\title{Bulk viscosity of a binary mixture: the role of the intra-species interaction}

\author{Gabriele Parisi}
\email{gabriele.parisi@dfa.unict.it}
\affiliation{Galileo Galilei Institute for Theoretical Physics, Largo Enrico Fermi 2, I-50125 Firenze, Italy}

\author{Vincenzo Nugara}
\email{vincenzo.nugara@phd.unict.it}
%\affiliation{Department of Physics and Astronomy, University of Catania, Via S. Sofia 64, I-95125 Catania, Italy}
\affiliation{INFN-Laboratori Nazionali del Sud, Via S. Sofia 62, I-95123 Catania, Italy}

\author{Shams Ul Arfeen}
\email{shams.arfeen@sse.habib.edu.pk}
\affiliation{Habib University, University Avenue, Karachi 75290, Pakistan}

\author{Salvatore Plumari}
\email{salvatore.plumari@dfa.unict.it}
\affiliation{INFN-Laboratori Nazionali del Sud, Via S. Sofia 62, I-95123 Catania, Italy}
\affiliation{Department of Physics and Astronomy, University of Catania, Via S. Sofia 64, I-95125 Catania, Italy}

\author{Vincenzo Greco}
\email{greco@lns.infn.it}
\affiliation{INFN-Laboratori Nazionali del Sud, Via S. Sofia 62, I-95123 Catania, Italy}
\affiliation{Department of Physics and Astronomy, University of Catania, Via S. Sofia 64, I-95125 Catania, Italy}

\begin{abstract}
The bulk viscosity $\zeta$ is a transport coefficient which is of central importance for various areas of modern physics. In particular, its determination for a fluid with two components
is challenging, since it involves a complex interplay of multiple microscopic processes that operate on different time scales. Within the Chapman-Enskog framework, based on a series expansion of the Boltzmann distribution function, many previous works have derived the 1$^{\text{st}}$ order result for the $\zeta$ of a mixture. However, such a result fails to reproduce relevant physical features of the system, especially when the masses of the two components are similar. In this work we improve the 1$^{\text{st}}$ order Chapman-Enskog result by deriving the $\zeta$ at the 2$^{\text{nd}}$ order in the expansion. We show that this improved formula encodes many physical properties that the 1$^{\text{st}}$ order result misses: under specific conditions, the 2$^{\text{nd}}$ order result can be qualitatively and quantitatively very different from the 1$^{\text{st}}$ order one. Moreover, this result is compared against the $\zeta$ evaluated within the Green-Kubo formalism, by means of a numerical solution of the Relativistic Boltzmann equation. The agreement with respect to this benchmark is significantly improved when moving from the 1$^{\text{st}}$ to the 2$^{\text{nd}}$ order CE result. 
\end{abstract}

\maketitle

\section{Introduction}

Relativistic fluid dynamics provides an effective macroscopic framework for describing strongly interacting matter under extreme conditions, such as those encountered in neutron stars and in ultra-relativistic heavy-ion collisions. In these systems, local thermodynamic equilibrium can be approximately established over sufficiently large spacetime scales, enabling a hydrodynamic description in terms of conserved currents and transport coefficients. Among the dissipative transport coefficients, the shear viscosity $\eta$ governs the response to transverse deformations, while the bulk viscosity $\zeta$ quantifies the response to radial ones.

Bulk viscosity, specifically, plays a quite important role across a wide range of physical systems, governing how matter responds to isotropic expansions or compressions, particularly when the masses of the constituent microscopic degrees of freedom are non-zero. In the Quark–Gluon Plasma (QGP) formed in heavy-ion collisions, it can significantly influence the hydrodynamic evolution near the QCD transition, where deviations from conformality are largest \cite{Karsch:2007jc, Meyer:2007dy, Ryu:2015vwa}. Moreover, it has been recently shown that some observables, such as the radial flow, are significantly sensitive to the value of the ratio $\zeta/s$ between bulk viscosity and entropy density \cite{Schenke:2020uqq}. In astrophysical environments, such as neutron stars, bulk viscosity is a key mechanism for damping density oscillations and for stabilizing oscillating modes (e.g. the so-called r-mode oscillations, restored by the Coriolis force), thereby affecting the rotational evolution and gravitational wave emission of the star \cite{Alford:2011pi, Jones:2001ya}. In cosmology, bulk viscous effects have been invoked to describe entropy production and effective pressure corrections in the early universe \cite{Weinberg:1971mx, Zimdahl:1996ka}. Across these contexts, bulk viscosity encodes essential information about microscopic interactions and departures from equilibrium, making it a crucial ingredient in the macroscopic description of relativistic fluids.

The evaluation of the bulk viscosity is particularly challenging
when the medium is composed of more than one species \cite{Gavassino:2023eoz,Hernandez:2025zxw,Sappi:2025tcx}. This is so because in a multi-component fluid one has to take into account both intra-species ($1\leftrightarrow1$ and $2\leftrightarrow2$) and inter-species ($1\leftrightarrow2$) interactions, each of which
may contribute differently to the transport properties. As a result, the system cannot be characterized by a single relaxation scale, and the bulk viscosity receives contributions from multiple dissipative channels. The bulk viscosity of binary mixtures can be calculated within
the Chapman–Enskog formalism, which provides a
systematic framework to extract transport coefficients
from the Boltzmann equation, based on an expansion of
the phase space distribution function of the fluid around the
equilibrium.\footnote{A similar issue has been tackled for the shear viscosity of a binary mixture, e.g. by some of the authors in \cite{Parisi:2025gwq}.} While the CE result for $\zeta$ of a binary fluid in the 1$^{\text{st}}$ order approximation is widely known in the literature \cite{Wiranata:2009cz,VANLEEUWEN1975233,Dash:2019zwq}, it has various limitations. Among these, the most notable is its inability to capture the physical features arising from intra-species interactions. In particular, it can be shown that in the homogeneous limit, in which all masses and cross sections are equal and so the two components become indistinguishable, the 1$^{\text{st}}$ order CE formula always gives $\zeta=0$. As a consequence, this relation is not trustworthy, especially when the masses of the two components are similar.

The main purpose of this work is to address this issue by evaluating the bulk viscosity at the 2$^{\text{nd}}$ order in the Chapman-Enskog approximation for a two-component mixture of relativistic classical gases, assuming only binary collisions in the collision term. As we will see, at this order the $\zeta$ captures the physical features deriving from both the intra-species and the inter-species interactions. Moreover, we will show that in the homogeneous limit our 2$^{\text{nd}}$ order formula correctly reduces to the $\zeta$ of a single fluid at 1$^{\text{st}}$ order. Finally, we will compare our findings with various results from numerical simulations of the relativistic Boltzmann equation in the Green-Kubo formalism, which makes use of the linear-response theory in order to relate the transport coefficients with the dissipation of fluctuations. We remark that our goal is the computation of a unique bulk viscosity $\zeta$ which is able to characterize the relaxation of the mixture as a whole. This means that, being $\Pi$ the bulk viscous pressure of the mixture, computed from the total energy-momentum tensor $T^{\mu\nu}$ and the equation of state of the medium, the $\zeta$ of the mixture satisfies the following relation:
\begin{equation}
    \Pi = -\zeta \,\nabla_\mu u^\mu,
\end{equation}
at late times, which is known as the Navier-Stokes limit. Here $\nabla_\mu$ is the 3-derivative orthogonal to the four-velocity $u^\mu$, identified through the Landau matching from $T^{\mu\nu}$ \cite{Denicol:2021clh}.
More generally, we aim to describe the system as a unique fluid whose transport coefficients take into account all the microscopic inter- and intra-species interactions.

The paper is structured as follows. In Section \ref{sec:CE} we introduce the Chapman-Enskog formalism for the bulk viscosity, both for homogeneous fluids (Sec. \ref{sec:1st_CE_homogeneous}) and for binary fluids at the 1$^{\text{st}}$ (Sec. \ref{sec:1st_CE_mixture}) and at the 2$^{\text{nd}}$ order (Sec. \ref{sec:2nd_CE_mixture}, which contains Eq. \eqref{eq:zeta_2comp_2order},  the main result of the paper) in the CE expansion. In Sec. \ref{sec:GK} we introduce the Green-Kubo method, which allows the evaluation of transport coefficients via numerical simulations of the Boltzmann equation. In Sec. \ref{sec:results} we present our findings, comparing the Green-Kubo and the Chapman-Enskog results for $\zeta$. Finally, in Sec. \ref{sec:concl} we summarize and conclude. This work includes three Appendices: these contain technical details on the analytical and numerical calculations of the paper, which have not been included in the main text for the sake of clarity. 

Throughout this work we will use natural units of measure, i.e. we assume $\hbar=c=k_B=1$.

\section{Bulk viscosity in Chapman-Enskog}
\label{sec:CE}
The Chapman-Enskog (CE) expansion is the conventional formalism used to derive fluid dynamic quantities, starting from the Boltzmann equation $p^\mu \partial_\mu f=\mathcal{C}[f]$. It aims to obtain a solution of the transport equation valid in the hydrodynamic stage of equilibration, i.e. in the phase in which the spatial non-uniformities slowly disappear and the system relaxes to complete equilibrium \cite{DeGroot:1980dk, Chapman_book_1974, Israel_1963}. The CE formalism corresponds to the microscopic implementation of an expansion in powers of the gradients, or equivalently of the Knudsen number, i.e. the ratio between the microscopic mean free path and the macroscopic (hydrodynamical) length scale. The linear term, which is the one under consideration in this paper, corresponds to linear laws for the transport phenomena.
%The series generated by the Chapman-Enskog procedure is generally thought to be asymptotic rather than convergent: this indicates that the first Chapman-Enskog approximation, which corresponds to linear laws for the transport phenomena, in general may not contain sufficient information on the underlying physical system.

In more detail, for a system not too far from equilibrium, one can express the Lorentz-invariant distribution function $f$ as follows \cite{DeGroot:1980dk, Wiranata:2012br, Prakash:1993bt}:
\begin{equation}
    f=f_0(1+\phi),
    \label{eq:distribution_function_CE}
\end{equation}
in which the function $\phi$ describes the 
deviation from the local equilibrium Boltzmann distribution function $f_0$ and it is assumed to satisfy $|\phi|\ll 1$, which allows us to neglect higher order powers of $\phi$. Then, the corrections to the local distribution function, included in the term $\phi$, can be expanded in terms of Laguerre polynomials depending on $E_k/T$, $E_k$ being the kinetic energy of the particles and $T$ the system temperature \cite{VANLEEUWEN197365, VANLEEUWEN1975233}. This enables the calculation of transport coefficients such as the shear viscosity $\eta$ and the bulk viscosity $\zeta$. This approach provides flexibility in choosing both the order of approximation and the type of medium to be studied, whether it is a homogeneous gas or a multi-component mixture.

\subsection{Chapman-Enskog for a homogeneous fluid}
\label{sec:1st_CE_homogeneous}

From the Chapman-Enskog expansion, one obtains the transport coefficients as an infinite sum, corresponding to the contribution of all the orders of the expansion. By following the approach outlined in \cite{VANLEEUWEN197365}, one can compute the bulk viscosity $\zeta$ within the CE approximation for the general case of relativistic particles with finite mass $m$ obeying Boltzmann statistics, colliding with a non-isotropic, energy-dependent differential cross section $\sigma=\sigma(s,t)$, being $s$ and $t$ the well-known Mandelstam variables, according to the collision integral:

\begin{align}
\mathcal C\left[f^i, f^j\right]_{\mathbf{p}} =&
\intop dP_2 \intop dP_{1'} \intop dP_{2'}\left(f^i_{1'}f^j_{2'}-f^i_{1}f^j_{2}\right) \nonumber\\ &\times \left|\mathcal{M}\right|^{2}\delta^{\left(4\right)}\left(p_{1}+p_{2}-p_{1'}-p_{2'}\right),
\end{align}
where $f^i_r=f^i(p_r)$, the subscript $i,j$ identify the different particle species and $dP=d^3p/((2\pi)^3 E_p)$.
$\mathcal{M}$ denotes the transition amplitude for the elastic processes {$|{\cal M}|^2=16 \pi\,{s}^2\, \sigma(s,t)$.} Note that the Chapman-Enskog approximation relies on elastic scattering only, hence in this model the particles retain their species after each collision and non-particle-conserving processes are not taken into account.\\

We start by reporting well known results for the homogeneous case. The first and second order approximations for the bulk viscosity of a homogeneous fluid are given by \cite{Wiranata:2012br, VANLEEUWEN197365}:

\begin{align}
    \zeta_{\text{hom}}^{(1)}&=\frac{\bar{\alpha}_2^2}{\bar{a}_{22}}T,\label{eq:zeta_1comp_1storder}\\
    \zeta_{\text{hom}}^{(2)}&=\frac{\bar{\alpha}_2^2 \bar{a}_{33}-2\bar{\alpha}_2 \bar{\alpha}_3 \bar{a}_{23}+\bar{\alpha}_3^2 \bar{a}_{22}}{\bar{a}_{22}\bar{a}_{33}-\bar{a}_{23}^2}T,\label{eq:zeta_1comp_2ndorder}
\end{align}
where:
\begin{align}
    \bar{\alpha}_2=&\frac32 \left[z\hat{h}\left(\gamma-\frac53 \right)+\gamma\right],\\
    \bar{\alpha}_3=&\hat{h}\left[\left(\frac32 z^2+\frac14 z\right)\left(\gamma-\frac53\right)-\frac56 z\right]\nonumber\\
    &-\frac12 z^2 \left(\gamma-\frac53 \right)+\gamma \left(\frac32 z+\frac{11}{4}\right),\\
    \bar{a}_{22}=&2\omega_{0}^{(2)},\label{eq:zeta_1comp_parameters}\\
    \bar{a}_{23}=&2z(\omega_{0}^{(2)}-\omega_{1}^{(2)})+\omega_{0}^{(2)},\\
    \bar{a}_{33}=&2z^2 (\omega_{0}^{(2)}-2\omega_{1}^{(2)}+\omega_{2}^{(2)})\nonumber\\
    &+z(2\omega_{0}^{(2)}-7\omega_{1}^{(2)})+\frac72 \omega_{0}^{(2)}.
\end{align}
In the above expressions $\hat{h}\equiv K_3(z)/K_2(z)$, being $K_n(z)$ the $n^{\text{th}}$ order modified Bessel function of the second kind with $z\equiv m/T$, while $\omega_i^{(s)}$ are the so-called relativistic omega integrals:
\begin{align}
    \omega_i^{(s)}&=\frac{2\pi z^3}{K_2(z)^2}\int_0^{+\infty}d \psi\, \sinh^7\psi\, \cosh^i\psi\nonumber\\
    &\cdot K_j(2z\cosh\psi)\int_0^\pi d \theta\, \sin\theta\,\sigma(\psi,\theta)\,(1-\cos^s\theta),
    \label{eq:single_component_omegaintegrals}
\end{align}
expressed in terms of
\begin{equation}
    \cosh\psi=\frac{\sqrt{s}}{2m},~~~j=\frac{5}{2}+\frac{1}{2}(-1)^i.
    \label{eq:single_component_omegaintegrals2}
\end{equation}
Finally, $\gamma$ is given by
\begin{equation}
    \gamma=\frac{c_p}{c_v},~~~~c_p=\frac{\dd\hat{h}(z)}{\dd T},~~~~c_v=c_p-\frac{1}{m}.
    \label{eq:single_component_gamma_cp_cv}
\end{equation}

\subsection{\texorpdfstring{$1^{\text{st}}$}{1} order Chapman-Enskog for binary mixtures}
\label{sec:1st_CE_mixture}

The Chapman-Enskog formalism also allows for the calculation of the bulk viscosity in binary mixtures. In these systems, the two species interact with distinct cross sections $\sigma_{11},\sigma_{22}$ among themselves and also with each other according to an inter-species cross section $\sigma_{12}$. While the formalism can, in principle, be generalized to mixtures with three or more components, such extensions introduce substantial algebraic complexity. For this reason, we restrict our analysis to the binary case, which already captures the essential features of multi-component systems within the CE formalism. This 2-component extension follows from a straightforward generalization of Eq. \eqref{eq:distribution_function_CE}, see \cite{DeGroot:1980dk,VANLEEUWEN1975233} for details.

Here we report results which are already well-known in the literature, see e.g. \cite{Wiranata:2009cz,VANLEEUWEN1975233,Dash:2019zwq}. We do so in order to introduce the main physical properties of the $\zeta$ of a binary mixture, and also to shed light on the limitations of the $1^{\text{st}}$ order CE approximation in the non-homogeneous case. Such limitations are solved by going to higher orders, as we will show.\\

At $1^{\text{st}}$ order in the CE approximation, the bulk viscosity of a binary mixture has the same formal expression as that for a one component gas, that is \cite{VANLEEUWEN1975233,Wiranata:2009cz}:
\begin{equation}
    \zeta_{\text{mix}}^{(1)}=T\frac{\alpha_2^{\,2}}{a_{22}},
    \label{eq:twocompeta_1storder}
\end{equation}
where now
\begin{align}
    \alpha_2&=x_1\frac{\gamma_{1}-\gamma}{\gamma_{1}-1}\label{eq:twocompzeta_1storder_alpha_2}\\
    a_{22}&=\frac{16 \rho_1 \rho_2}{M_{12}^2 n^2}\omega_{1200}^{(1)}(\sigma_{12}).\label{eq:twocompzeta_1storder_a_22}
\end{align}

In the above relations $x_i=n_i/n$ is the number fraction of the $i$-th component, where $n_i$ is the particle number density of particle type $i$, and $n=n_1+n_2$ is the total particle number density. Moreover $\rho_i=m_in_i$ is the mass density of each species and $\rho=\rho_1+\rho_2$ the total mass density. We have labeled $M_{12}=m_1+m_2$ the sum of the two masses. The expression for the generalized relativistic omega integrals $\omega_{rtuv}^{(s)}$ is:
\begin{align}\omega_{rtuv}^{(s)}=&\frac{\pi \mu_{12}}{4T K_2(z_1)K_2(z_2)}\int_0^{+\infty}d \Psi_{12}\sinh^3\Psi_{12}\nonumber\\
&\cdot\left(\frac{g_{12}^2}{2\mu_{12}T}\right)^r\left(\frac{M_{12}}{P_{12}}\right)^t(\cosh \psi_1)^u (\cosh \psi_2)^v\nonumber \\
&\cdot K_\nu \left(\frac{P_{12}}{T}\right)\int_0^\pi d\theta \sin \theta \,\sigma(\Psi_{12},\theta)(1-\cos^s\theta),
\label{eq:omega_rtuv}
\end{align}
where:
\begin{gather}
\Psi_{12}\equiv \psi_1+\psi_2,~~~\nu=\frac{5}{2}-\frac{1}{2}(-1)^{t+u+v},\nonumber\\
P_{12}^2=m_1^2+m_2^2+2m_1m_2\cosh\Psi_{12},\nonumber\\
g_{12}=\frac{m_1m_2\sinh\Psi_{12}}{P_{12}},\nonumber\\
\cosh\psi_1=\frac{1}{P_{12}}(m_1+m_2\cosh\Psi_{12}),\nonumber\\
\cosh\psi_2=\frac{1}{P_{12}}(m_2+m_1\cosh\Psi_{12}),\nonumber\\
\mu_{12}=m_1 m_2/M_{12} \label{Psi_Ptot_ecc}.
\end{gather}
Above $P_{12}=\sqrt{-(p_1+p_2)^2}$ is the invariant center-of-mass energy of the two particles colliding with initial four momenta $p_1$ and $p_2$. Finally, the ratios $\gamma_1$ and $\gamma$ appearing in \eqref{eq:twocompzeta_1storder_alpha_2} are given by:
\begin{equation}
\gamma_{i}=\frac{c_p^{(i)}}{c_v^{(i)}},~~~~c_p^{(i)}=\frac{\dd\hat{h}(z_i)}{\dd T},~~~~c_v^{(i)}=c_p^{(i)}-\frac{1}{m_i},
    \label{eq:twocompzeta_1storder_gamma_gamma1}
\end{equation}
\begin{equation}
    \gamma=\frac{c_p}{c_v},~~~~c_p=c_1c_p^{(1)}+c_2c_p^{(2)},~~~~c_v=c_1c_v^{(1)}+c_2c_v^{(2)},
    \label{eq:twocompzeta_gamma_cp_cv}
\end{equation}
where $z_i\equiv m_i/T$ and $c_i$ is the mass fraction of the $i$-th component, defined as $c_i=\rho_i/\rho$. Notice that \eqref{eq:twocompzeta_1storder_alpha_2} is not explicitly symmetric with respect to the two components $1\leftrightarrow 2$. The asymmetry is however only apparent, since it is easy to show, using \eqref{eq:twocompzeta_1storder_gamma_gamma1} and \eqref{eq:twocompzeta_gamma_cp_cv}, that
\begin{equation}
x_1\frac{\gamma_{1}-\gamma}{\gamma_{1}-1}=-x_2\frac{\gamma_{2}-\gamma}{\gamma_{2}-1},
    \label{eq:single_component_alpha2_asymmetry}
\end{equation}
so the bulk viscosity $\zeta_{\text{mix}}^{(1)}$ in Eq. \eqref{eq:twocompeta_1storder} is rightfully symmetric with respect to the swap of indices $1\leftrightarrow 2$.

\subsection{\texorpdfstring{$2^{\text{nd}}$}{2} order Chapman-Enskog for binary mixtures}
\label{sec:2nd_CE_mixture}

At this point let us note that, intuitively, if we consider a binary mixture in which both masses are equal ($m_1=m_2=m$) and all the cross sections are equal as well ($\sigma_{11}=\sigma_{22}=\sigma_{12}=\sigma$), what we get is a homogeneous gas, a single-component fluid whose particles have mass $m$ and interact with a differential cross section $\sigma$. The CE formalism for a binary mixture should reproduce this result in the aforementioned limits. However, if we consider both masses equal we immediately see that $\gamma_1=\gamma_2=\gamma$, and therefore from Eqs. \eqref{eq:twocompeta_1storder} and \eqref{eq:twocompzeta_1storder_alpha_2} we get $\zeta_{\text{mix}}^{(1)}=0$. Moreover, we also see that $\zeta_{\text{mix}}^{(1)}$ does not depend in any way on the intra-species cross sections $\sigma_{11}$ and $\sigma_{22}$. These two observations bring us to the conclusion that the $1^{\text{st}}$ order Chapman-Enskog approximation does not capture all the physical features of the bulk viscosity of a binary mixture.

In order to shed light on this point let us recall that, in a homogeneous mixture, while the shear viscosity contributes to the damping of both shear and sound modes, the bulk viscosity contributes to the damping of the sound modes only (in the expression of the viscous stress tensor $\Pi_{\mu\nu}$, $\zeta$ is proportional to the divergence of the fluid velocity). On the other hand, in a binary mixture, we have the presence of additional hydrodynamic modes related to concentration diffusion, which originate from the conservation of the individual species' densities. 
Since in this case the main contribution to the bulk viscosity is given by the relaxation of these inter-species modes, its leading term will be controlled by collisions between dissimilar species, i.e. by the inter-species cross section only \cite{Dash:2019zwq}. Note that this does not occur when one employs the Chapman-Enskog expansion to evaluate the shear viscosity of a binary mixture, see \cite{Wiranata:2013oaa, Parisi:2025gwq}, since these extra hydrodynamic modes are excited by compressional (longitudinal) perturbations and therefore they couple to bulk viscosity, but not to shear deformation.

In light of these arguments, we understand that when we perform the homogeneous limit on the $1^{\text{st}}$ order formula for the bulk viscosity Eq. \eqref{eq:twocompeta_1storder}, we correctly obtain zero, since there is no concentration diffusion in a homogeneous fluid and $\zeta_{\text{mix}}^{(1)}$ captures only this physics. While in a binary mixture both particle numbers are conserved (we have elastic collisions only), this is not so for the individual energy densities. Indeed, the energy density of the individual species changes as a result of the inter-species interactions, and only the total energy is conserved. As a result, in a binary mixture there are less conservation laws to be enforced, hence leading to non-zero results.
Collisions among identical species are of course present, in general, and one expects their effect to come up at higher orders. The previous considerations justify and lead to the main result of this paper, that is the explicit derivation of the bulk viscosity $\zeta_{\text{mix}}^{(2)}$ for a binary mixture at the second order in the CE approximation.\\

Here we limit ourselves to report the final expression only, which has been developed starting from the results of \cite{VANLEEUWEN1975233}. More details can be found in Appendix \ref{sec:appendix_details_of_calculations}. In addition to that, in Appendix \ref{sec:appendix_reduction2to1comp} we show that when we perform the homogeneous limit on our result $\zeta_{\text{mix}}^{(2)}$, we now obtain the $1$-component formula \eqref{eq:zeta_1comp_1storder} instead of the trivial result $\zeta=0$. The full expression for $\zeta_{\text{mix}}^{(2)}$ is

\onecolumngrid

\begin{align}
    \zeta_{\text{mix}}^{(2)}=&\frac{1}{\mathcal{D}}\left[(a_{2,3}^2-a_{2,2}\,a_{3,3})\alpha_{-3}^2+2(a_{2,2}\,a_{3,-3}+a_{2,3}\,a_{-2,-3})\alpha_{-3}\,\alpha_3+(a_{-2,-3}^2-a_{2,2}\,a_{-3,-3})\alpha_3^2\right.\nonumber\\
    &\left.+(a_{3,-3}^2-a_{3,3}\,a_{-3,-3})\alpha_2^2-2(a_{2,3}\,a_{3,-3}+a_{3,3}\,a_{-2,-3})\alpha_2\, \alpha_{-3}+2(a_{-2,-3}\,a_{3,-3}+a_{-3,-3}\,a_{2,3})\alpha_2 \,\alpha_{3}\right]T,
    \label{eq:zeta_2comp_2order}
\end{align}
where the denominator $\mathcal{D}$ is
\begin{equation}
    \mathcal{D}\equiv a_{2,2}(a_{3,-3}^2-a_{3,3}\,a_{-3,-3})+a_{2,3}^2\, a_{-3,-3}+a_{-2,-3}^2\,a_{3,3}+2a_{2,3}\,a_{-2,-3}\,a_{3,-3}.
    \label{eq:denominator_D}
\end{equation}
The expressions for the coefficients $\alpha_r$ and $a_{r,s}$ are the following:
\begin{align}
    \alpha_2=&x_1\frac{\gamma_{1}-\gamma}{\gamma_{1}-1},\label{eq:zeta_2ndorder_alpha2}\\
    \alpha_3=&x_1\left\{(z_1+1)\frac{\gamma_{1}-\gamma}{\gamma_{1}-1}+\frac32\left[z_1\hat{h}_1\left(\gamma-\frac53\right)+\gamma\right]\right\},\label{eq:zeta_2ndorder_alpha3}\\
    \alpha_{-3}=&x_2\left\{(z_2+1)\frac{\gamma_{2}-\gamma}{\gamma_{2}-1}+\frac32\left[z_2\hat{h}_2\left(\gamma-\frac53\right)+\gamma\right]\right\},\label{eq:zeta_2ndorder_alpha-3}
    \end{align}
    \begin{align}
    a_{2,2}=&\frac{16 \rho_1 \rho_2}{M_{12}^2 n^2}\omega_{1200}^{(1)}(\sigma_{12}),\label{eq:zeta_2ndorder_a22}\\
    a_{2,3}=&\frac{4 \rho_1 \rho_2}{M_{12}^2 n^2}\left[(10+4z_1)\,\omega_{1200}^{(1)}(\sigma_{12})-4z_1\,\omega_{1210}^{(1)}(\sigma_{12})\right],\label{eq:zeta_2ndorder_a23}\\
    a_{-2,-3}=&\frac{4 \rho_1 \rho_2}{M_{12}^2 n^2}\left[(10+4z_2)\,\omega_{1200}^{(1)}(\sigma_{12})-4z_2\,\omega_{1201}^{(1)}(\sigma_{12})\right],\label{eq:zeta_2ndorder_a-2-3}
\end{align}
\begin{align}
    a_{3,3}=&2x_1^2\,\omega_{0}^{(2)}(m_1,\sigma_{11})+\frac{4\rho_1 \rho_2}{M_{12}^2 n^2}\left[4\frac{\mu_{12}}{M_{12}}\omega_{2300}^{(2)}(\sigma_{12})\right.\nonumber\\
    &\left.+(5+2z_1)^2\omega_{1200}^{(1)}(\sigma_{12})+4z_1^2\, \omega_{1220}^{(1)}(\sigma_{12})+10\frac{z_1^2}{Z_{12}}\omega_{1320}^{(1)}(\sigma_{12})-4z_1(5+2z_1)\omega_{1210}^{(1)}(\sigma_{12})\right],\label{eq:zeta_2ndorder_a33}\\
    a_{-3,-3}=&2x_2^2\,\omega_{0}^{(2)}(m_2,\sigma_{22})+\frac{4\rho_1 \rho_2}{M_{12}^2 n^2}\left[4\frac{\mu_{12}}{M_{12}}\omega_{2300}^{(2)}(\sigma_{12})\right.\nonumber\\
    &\left.+(5+2z_2)^2\omega_{1200}^{(1)}(\sigma_{12})+4z_2^2\, \omega_{1202}^{(1)}(\sigma_{12})+10\frac{z_2^2}{Z_{12}}\omega_{1302}^{(1)}(\sigma_{12})-4z_2(5+2z_2)\omega_{1201}^{(1)}(\sigma_{12})\right],\label{eq:zeta_2ndorder_a-3-3}\\
    a_{3,-3}=&\frac{4\rho_1 \rho_2}{M_{12}^2 n^2}\left[4\frac{\mu_{12}}{M_{12}}\omega_{2300}^{(2)}(\sigma_{12})-25\omega_{1200}^{(1)}(\sigma_{12})-10(z_1+z_2)\omega_{1200}^{(1)}(\sigma_{12})-4z_1z_2 \omega_{1200}^{(1)}(\sigma_{12})\right.\nonumber\\
    &-\left.4 z_1 z_2 \omega_{1211}^{(1)}(\sigma_{12})-10\frac{z_1z_2}{Z_{12}}\omega_{1311}^{(1)}(\sigma_{12})+2z_2(5+2z_1)\omega_{1201}^{(1)}(\sigma_{12})
    +2z_1(5+2z_2)\omega_{1210}^{(1)}(\sigma_{12})\right].\label{eq:zeta_2ndorder_a3-3}
\end{align}
In the above relations we have introduced $Z_{12}=M_{12}/(2T)$. 
\vspace{10pt}
\twocolumngrid

\section{Bulk viscosity in Green-Kubo}
\label{sec:GK}
From the kinetic theory perspective, the bulk viscosity $\zeta$ can be put in relation to the correlation function of the bulk viscous pressure, as done also for other transport coefficients and their corresponding fluxes or tensors at thermal equilibrium \cite{Green:1954ubq,Kubo_1957}. The fluctuation-dissipation theorem ensures that dissipation of fluctuations and relaxation towards equilibrium have the same physical origin: both are dominated by the same transport coefficients \cite{Wiranata:2012br}. Hence, it is  possible to derive them from the microscopic model, by using linear response theory, within the so-called Green-Kubo (GK) method. This approach has successfully been employed in the literature, also by some of the authors, to investigate the shear viscosity of single-component \cite{Plumari:2012ep} and two-component \cite{Parisi:2025gwq} gases. Some results have also been obtained for the bulk viscosity of a one-component gas \cite{Rose:2020lfc}, while this is, to our knowledge, the first time the bulk viscosity of a mixture is studied. The Green-Kubo formalism is indeed very versatile and easily allows generalizations, such as the one here considered, without any need to modify the approach. Therefore, we use the Green-Kubo results as a benchmark to validate our second-order Chapman-Enskog formula for the bulk viscosity of a binary mixture.

According to the Green-Kubo method, the expression of the bulk viscosity for a medium at fixed temperature is given by the formula \cite{zubarev1996statistical}:
\begin{equation}
    \zeta=\beta\lim_{T_{\text{max}}\to +\infty}\int_0^{T_{\text{max}}}\dd t \int_V \dd^3 \mathbf{x}\, \langle \delta\Pi(\mathbf{x},t)\delta\Pi(\mathbf{x},0)\rangle,
    \label{4.1}
\end{equation}
where $\beta$ is the inverse of the temperature and ${\delta\Pi}=\Pi - \bar \Pi$ gives the fluctuation of the bulk viscous pressure $\Pi$ with respect to its value averaged over total time $\bar \Pi$. Here $\langle ... \rangle$ denotes the following convolution procedure:
\begin{align}
\left\langle \delta\Pi(\mathbf{x},t)\right.&\left.\delta\Pi(\mathbf{x},0) \right\rangle =\nonumber \\
&\lim_{\mathcal{T}_{\text{max}} \to +\infty} \frac{1}{\mathcal{T}_{\text{max}}} \int_0^{\mathcal{T}_{\text{max}}} dt' \, \delta\Pi(\mathbf{x},t + t')\delta\Pi(\mathbf{x},t'). \label{eq:pi_pi_time_correlator}
\end{align}

In this paper, the bulk pressure fluctuation correlator $\langle \delta\Pi(\mathbf{x},t)\delta\Pi(\mathbf{x},0)\rangle$ is computed through transport simulations of a system at thermal equilibrium. Such a system is confined in a static box of volume $V$ and subject to periodic boundary conditions, similarly to what has been done in \cite{Plumari:2012ep, Parisi:2025gwq}. These simulations are performed employing a relativistic transport code developed, in recent years, to describe the Quark-Gluon Plasma dynamics of heavy-ion collisions for different collision systems \cite{Ferini:2008he, Scardina:2012mik, Puglisi:2014sha, Plumari:2015cfa, Scardina:2017ipo, Plumari:2019hzp, Sun:2019gxg, Sambataro:2020pge, Sambataro:2023tlv, Oliva:2020doe, Nugara:2024net}.
The purpose of the code is to numerically solve the relativistic Boltzmann transport equation with the full elastic collision integral. This is possible by means of the stochastic algorithm \cite{Xu:2004mz}, i.e. a  Monte-Carlo numerical method that allows us to treat in a Lorentz-covariant and causality-preserving fashion the full collision kernel, mapping the evolution of the distribution function in discretized time. Thus, we use the solution of the kinetic equation to extract the bulk viscosity using the Green-Kubo formula, and compare the results with those obtained within the Chapman-Enskog formalism. 
The version employed herein is optimized for box calculations, such as the ones needed in this work, allowing a very fast estimation of correlators. More details on the numerical implementation of the Green-Kubo method are available in Appendix \ref{sec:appendix_numerical}.

We now turn to the computation of the bulk viscous component of the energy-momentum tensor. Since there is no spatial inhomogeneity (e.g. preferred directions or external fields), we consider the volume-averaged bulk pressure $\Pi(t)$, as well as its time average $\bar \Pi(t)$ and fluctuations $\delta\Pi(t)$. The evaluation of $\Pi(t)$ is performed as
\begin{equation}
\Pi(t)= \frac{T_{xx} + T_{yy} + T_{zz}}{3} - P_{eq}  = \frac{1}{3V}\sum_{k=1}^{N} \frac{p_k^2}{E_k} -\frac{N}{V} T,
    \label{eq:discretized_pi_xy}
\end{equation}
where $N$ is the total number of test particles used and $P_{eq}=nT$. We finally derive the bulk viscosity by discretizing Eq. \eqref{4.1}:
\begin{equation}
\zeta=\beta\,V \Delta t\sum_{j=0}^{N_{T_{\text{max}}}}\left\langle\delta\Pi(j\Delta t) \delta\Pi(0)\right\rangle,
    \label{4.1_discretized}
\end{equation}
{where $N_{T_{\text{max}}}=T_{\text{max}}/\Delta t\approx 500$, $T_{\text{max}}$ is the final time chosen for the correlator} and the discretization of the convolution procedure $\langle ...\rangle$ in \eqref{eq:pi_pi_time_correlator} has been performed as
\begin{align}
\left\langle \delta\Pi(j\Delta t)\right.&\left.\delta\Pi(0) \right\rangle=\nonumber\\
&\frac{1}{N_{\mathcal{T}_{\text{max}}}} \sum_{k=0}^{N_{\mathcal{T}_{\text{max}}}} \delta\Pi(j\Delta t + k\Delta t)\delta\Pi(k\Delta t)\label{eq:pi_pi_time_correlator_discretized}
\end{align}
where $N_{\mathcal{T}_{\text{max}}}=\mathcal{T}_{\text{max}}/\Delta t\approx 30\,000$ ($\Delta t$ is the same for both \eqref{4.1_discretized} and \eqref{eq:pi_pi_time_correlator_discretized}). In order to have a good accuracy, $\mathcal{T}_{\text{max}}$ has to be chosen significantly larger than $T_{\text{max}}$. In particular, for each event we fix $\mathcal{T}_{\text{max}}/T_{\text{max}}\approx 30\,000/500\approx 10^{2}$. One should remark that this large ratio also leads to a faster numerical convergence of the correlator, since the distribution function will be able to explore a broader region of the phase space in a longer time, leading to a smoother correlator. The numerical computation of the integral \eqref{4.1} is performed via the trapezoid method.

The robustness of the method which we have just illustrated can be cross-checked as follows. Since we expect the functional dependence of the correlator with respect to time to be a decreasing exponential, we can fit the discretized set of data with a function
\begin{equation}
    f(t)=\langle\delta\Pi(0)\delta\Pi(0)\rangle_\text{fit}\exp(-t/\tau_\text{fit}),
    \label{eq:fit_function_correlator}
\end{equation}
and, by extracting both $\langle\delta\Pi(0)\delta\Pi(0)\rangle_\text{fit}$ and $\tau_\text{fit}$, the evaluation of the bulk viscosity can be performed by analytically integrating the fitted function:
\begin{equation}
\zeta = \frac{V}{T} \langle\delta\Pi(0)\delta\Pi(0)\rangle_\text{fit}\: \tau_\text{fit}.
    \label{eq:zeta_fit_function_correlator}
\end{equation}
We find that this result is, within one standard deviation, in agreement with the value obtained by explicit integration, which validates the use of \eqref{4.1_discretized} for the evaluation of the bulk viscosity.

%For illustrative purposes, in Figure \ref{average_correlator_qpm_T_0.5} we show (in cyan) the result of 45 different calculations for the time behavior of $\langle \pi^{xy}(t) \pi^{xy}(0)\rangle $ at a temperature of $T=0.5$ GeV in the \textit{QPM, anisotropic} case (check Appendices \ref{Appendix B} and \ref{Appendix E} for details). Together with the results for each event, which carry a significant amount of fluctuations, we show their ensemble average (dark blue), which highlights the typical exponential decay of such correlator.\footnote{The shear viscosity can also be obtained, as in \cite{Plumari:2012ep}, by assuming an exponential decay for the correlator, and then by linking the proper $\tau$ of such decay to $\eta$ using the Green-Kubo formula \eqref{4.1}.} Notice that the negative values of the correlators are only a matter of numerical fluctuations. Indeed, they basically disappear in the averaged correlator and would be suppressed by further increasing the number of test particles.

\section{Results}
\label{sec:results}

\begin{figure}[ht]
    \centering
    \includegraphics[width=\linewidth]{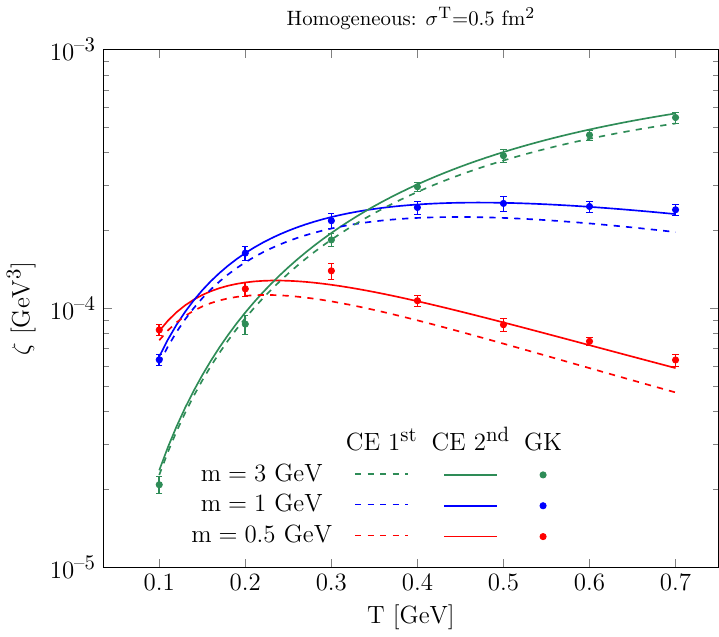}
    \caption{Bulk viscosity $\zeta$ of a homogeneous gas as a function of temperature, for various values of the mass $m$. Here we fix $\sigma^{\text{T}}$=0.5 fm$^2$.}
    \label{fig:Bulk_viscosity_CEvsGK_vs_temperature_1component}
\end{figure}

\begin{figure*}[ht]
    \centering
    \includegraphics[width=0.48\linewidth]{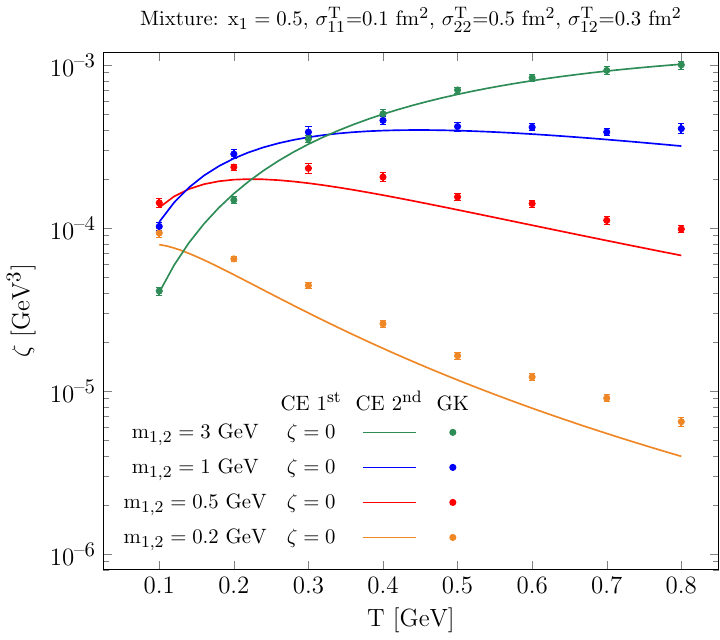}
    \includegraphics[width=0.48\linewidth]{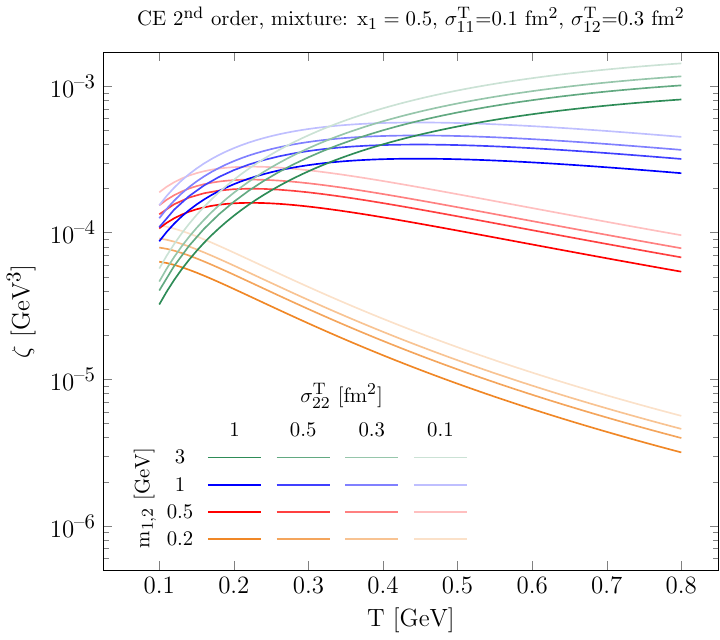}
    \caption{Bulk viscosity $\zeta$ of a binary mixture as a function of temperature. Left panel: we compare the 2$^{\text{nd}}$ order Chapman-Enskog with the Green-Kubo calculations, for various values of the two masses (here assumed equal). We fix the cross sections as $\sigma_{11}^{\text{T}}$=0.1 fm$^2$, $\sigma_{22}^{\text{T}}$=0.5 fm$^2$, $\sigma_{12}^{\text{T}}$=0.3 fm$^2$. Right panel: focusing on the 2$^{\text{nd}}$ order CE results only, for each value of the mass we show the results obtained by varying $\sigma_{22}^{\text{T}}$. Here $\sigma_{11}^{\text{T}}$=0.1 fm$^2$ and $\sigma_{12}^{\text{T}}$=0.3 fm$^2$ are kept fixed. In both plots the two components have the same number concentration, in other words we fix $x_1=x_2=0.5$.}
    \label{fig:Bulk_viscosity_CEvsGK_vs_temperature}
\end{figure*}

In this Section we compare the results obtained for $\zeta$ within the Green-Kubo approach, i.e. Eq. \eqref{4.1_discretized}, with the Chapman-Enskog estimates. We will start with the results for homogeneous fluids (CE Eqs. \eqref{eq:zeta_1comp_1storder} and \eqref{eq:zeta_1comp_2ndorder} at 1$^{\text{st}}$ and 2$^{\text{nd}}$ order, respectively), and then move on to binary mixtures (CE Eqs. \eqref{eq:twocompeta_1storder} and \eqref{eq:zeta_2comp_2order} at 1$^{\text{st}}$ and 2$^{\text{nd}}$ order, respectively). In order to highlight the core aspects of our findings, in the following we will ignore additional complications by considering only isotropic scatterings for both the GK and the CE calculations. Under this assumption, the total cross section and the differential cross section are simply related by $\sigma^{\text{T}}=4\pi \sigma$. In all the plots that follow, the total number density $n$ has been kept fixed: notice that our results are sensitive only to the relative density concentration $n_i/n\equiv x_i$, not to the total density $n$ only.\\

We start by studying the bulk viscosity of a homogeneous gas. In Figure \ref{fig:Bulk_viscosity_CEvsGK_vs_temperature_1component} we plot the bulk viscosity as a function of temperature for three different mass values (different colors) according to the 1$^\text{st}$ and 2$^\text{nd}$ order Chapman-Enskog formulae (respectively dashed and solid lines) and compare them with the Green-Kubo calculations (points). The values of the masses here are chosen in the same range as the ones employed in phenomenological quasi particle models (see e.g. \cite{Sambataro:2024mkr} and works cited therein). The interaction cross section is fixed as $\sigma^\text{T}=0.5$ fm$^2$. One clearly sees that, for the explored range of temperatures and masses, the two frameworks agree very well if one considers the 2$^\text{nd}$ order CE. Quite interestingly, the $y$ log scale highlights that the relative differences between GK and 1$^\text{st}$ order CE (or equivalently 2$^\text{nd}$ order CE) decrease with higher values of the mass. In the case $m=0.5$ GeV, we see that the 2$^\text{nd}$ order CE significantly improves the 1$^\text{st}$ order result, bringing the relative difference with respect to the GK calculations from $\sim 20\%$ down to $\sim$ 5$\%$ for the highest $T$. On the other hand, in the $m=3$ GeV case the 1$^\text{st}$ order CE already shows really good agreement for the whole range of $T$.\\

\begin{figure}[ht]
    \centering
    \includegraphics[width=0.96\linewidth]{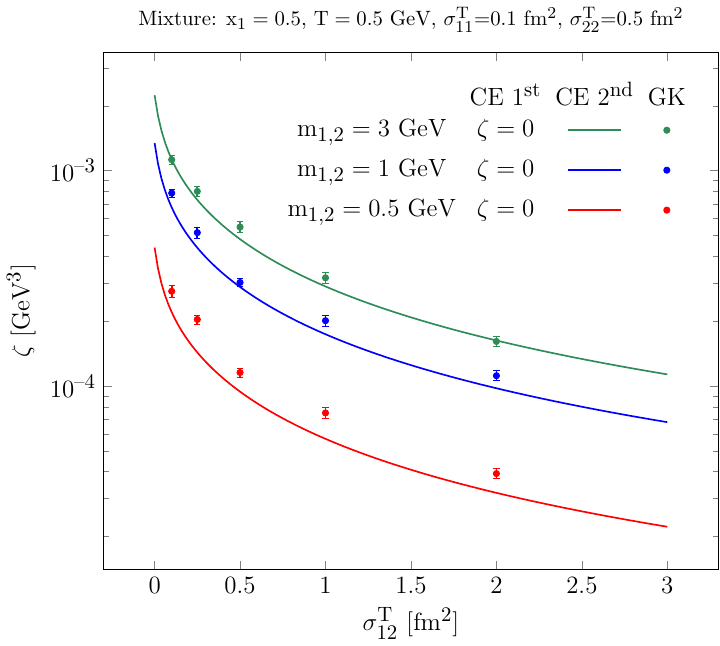}
    \caption{Bulk viscosity $\zeta$ of a binary mixture as a function of the inter-species total cross section $\sigma_{12}^{\text{T}}$, for various values of the two masses (here assumed equal). The other parameters have been fixed as follows: $x_1=0.5$, $T=0.5$ GeV, $\sigma_{11}^{\text{T}}$=0.1 fm$^2$, $\sigma_{22}^{\text{T}}$=0.5 fm$^2$.}
    \label{fig:Bulk_viscosity_CEvsGK_vs_sigma12}
\end{figure}

\begin{figure}[ht]
    \centering
    \includegraphics[width=0.96\linewidth]{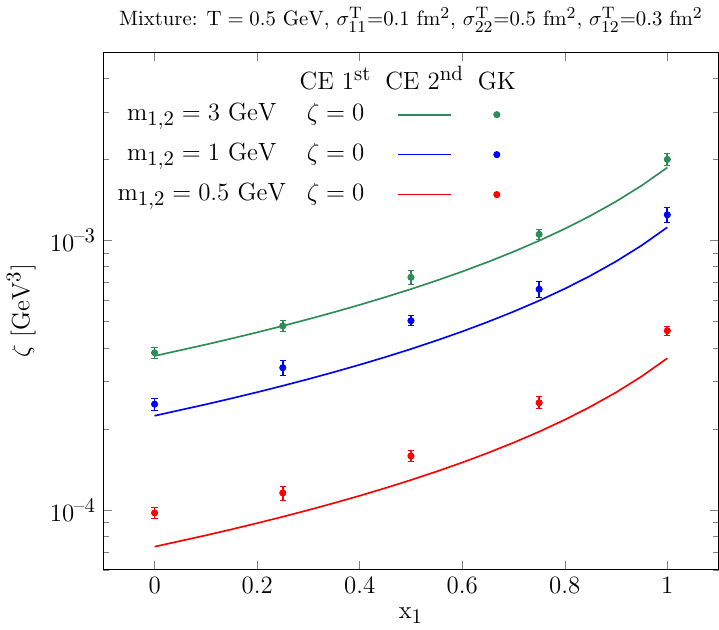}
    \caption{Bulk viscosity $\zeta$ of a binary mixture as a function of the number concentration $x_1$ of the first species, for various values of the two masses (here assumed equal). The other parameters have been fixed as follows: $T=0.5$ GeV, $\sigma_{11}^{\text{T}}$=0.1 fm$^2$, $\sigma_{22}^{\text{T}}$=0.5 fm$^2$, $\sigma_{12}^{\text{T}}$=0.3 fm$^2$.}
    \label{fig:Bulk_viscosity_CEvsGK_vs_concentration}
\end{figure}

Now we turn to the main focus of this work, that is the binary mixture. In Figures \ref{fig:Bulk_viscosity_CEvsGK_vs_temperature}, \ref{fig:Bulk_viscosity_CEvsGK_vs_sigma12} and \ref{fig:Bulk_viscosity_CEvsGK_vs_concentration} we show our results for the bulk viscosity $\zeta$ for various values of the masses $m_1=m_2$: in these plots the masses of the two species are equal, and the different species are distinguished by the different intra-species interaction cross sections $\sigma_{11}\ne\sigma_{22}$. 
Here we show the curves obtained for $m_1=m_2$ in order to highlight the distinctive features that the 1$^{\text{st}}$ order CE is not able to describe. In particular, we show the dependence of $\zeta$ with respect to temperature $T$ (Figure \ref{fig:Bulk_viscosity_CEvsGK_vs_temperature}), with respect to the inter-species total cross section $\sigma_{12}^{\text{T}}$ (Figure \ref{fig:Bulk_viscosity_CEvsGK_vs_sigma12}) and with respect to the number concentration $x_1$ of the species 1 (Figure \ref{fig:Bulk_viscosity_CEvsGK_vs_concentration}). The specific choices of the other parameters which, plot by plot, have been kept fixed, are reported in the respective titles and captions.

As extensively argued, the first striking feature that these results have in common is that for equal masses the 1$^{\text{st}}$ order CE always gives $\zeta=0$, and therefore the corresponding lines do not appear in the plots at all. In addition to that, we see that the larger the common mass of the two components, the better the agreement between the 2$^{\text{nd}}$ order CE and the GK results, exactly as we have already seen in the case of the homogeneous gas. This leads us to state that for higher masses the CE series converges faster than it does for small masses, and we can attribute the difference to the absence of the 3$^{\text{rd}}$ and highest terms in the CE expansion.

In the left panel of Figure \ref{fig:Bulk_viscosity_CEvsGK_vs_temperature} we can see how the qualitative behavior of $\zeta$ vs $T$ changes with respect to a change in the mass of the components. In particular, while for $m_1=m_2=0.2$ GeV we see that $\zeta$ is a monotonically decreasing function of $T$, by increasing the mass we see a change of the shape of the curves, until we get a monotonically increasing function of $T$ for the highest value of mass. This can be understood by noting that, at fixed temperature, the ratio $z=m/T$ is larger for larger $m$. This means that for the highest values of mass we are further away from the conformal limit, in which $\zeta\to 0$, and we would need very high temperatures (not shown in the plot) to achieve $z\ll1$ and therefore a decreasing behavior of $\zeta$ towards zero. For the same reason, conversely, the decreasing trend with respect to $T$ appears sooner for the lowest values of mass. In the right panel of Figure \ref{fig:Bulk_viscosity_CEvsGK_vs_temperature} we focus on the 2$^{\text{nd}}$ order CE results only, to highlight the effect that a change in one of the intra-species cross sections has on the bulk viscosity: indeed, by increasing $\sigma_{22}^{\text{T}}$ we see that $\zeta$ decreases, regardless of the masses of the two components. Moreover, the qualitative behavior of $\zeta$ vs $T$ is the same within each set of curves with same mass (i.e. same colour). In Figure \ref{fig:Bulk_viscosity_CEvsGK_vs_sigma12} we highlight the behavior of $\zeta$ with respect to $\sigma_{12}^{\text{T}}$, observing that the $\zeta$ of a mixture strongly decreases with respect to an increase of the inter-species cross section. Notice that the dependence of $\zeta$ with respect to $\sigma_{12}^{\text{T}}$ is much stronger than with respect to $\sigma_{22}^{\text{T}}$ (we expect a similar conclusion for $\sigma_{11}^{\text{T}}$ as well). Indeed, a comparison between the right panel of Figure \ref{fig:Bulk_viscosity_CEvsGK_vs_temperature} and Figure \ref{fig:Bulk_viscosity_CEvsGK_vs_sigma12} shows that an increase of $\sigma_{22}^{\text{T}}$ by a factor $10$ reduces $\zeta$ by a factor smaller than 2, while the same reduction is achieved by only doubling $\sigma_{12}^{\text{T}}$ from $1$ to $2$ fm$^2$. In Figure \ref{fig:Bulk_viscosity_CEvsGK_vs_concentration} we see that $\zeta$ is an increasing function of $x_1$, that is, the bulk viscosity of a mixture increases when the concentration of the fluid with smaller cross section increases.

\begin{figure}[ht]
    \centering
    \includegraphics[width=\linewidth]{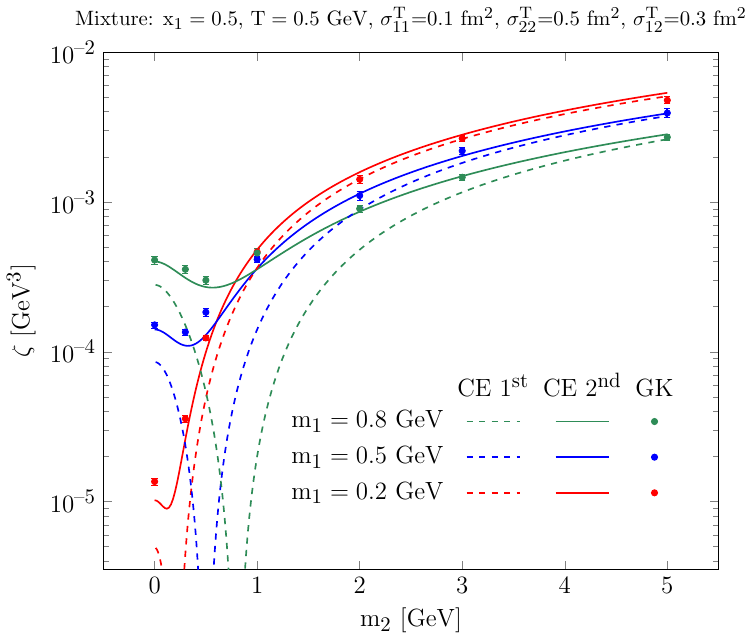}
    \caption{Bulk viscosity $\zeta$ of a binary mixture as a function of the mass $m_2$ of the second species, for various values of the mass $m_1$ of the first species. The other parameters have been fixed as follows: $x_1=0.5$, $T=0.5$ GeV, $\sigma_{11}^{\text{T}}$=0.1 fm$^2$, $\sigma_{22}^{\text{T}}$=0.5 fm$^2$, $\sigma_{12}^{\text{T}}$=0.3 fm$^2$.}
    \label{fig:Bulk_viscosity_CEvsGK_vs_mass2}
\end{figure}

Finally, in Figure \ref{fig:Bulk_viscosity_CEvsGK_vs_mass2} we lift the assumption of equal masses, and show $\zeta$ as a function of the mass $m_2$ of the second component, for various masses $m_1$ of the first component. The other parameters are kept fixed as in the title and in the caption. Despite in this case we can now appreciate the 1$^{\text{st}}$ order CE results being in general non-zero, each 1$^{\text{st}}$ order curve (dashed lines) lies below each corresponding 2$^{\text{nd}}$ order curve (full lines). In particular, the discrepancy is larger the more $m_2$ gets close to $m_1$, since in this case  $\zeta_{\text{mix}}^{(1)}\to 0$ and the discrepancy is maximal. Since in our formalism the species which we label as ‘2' is the one with largest intra-species cross section, from Figure \ref{fig:Bulk_viscosity_CEvsGK_vs_mass2} we derive the conclusion that the 1$^{\text{st}}$ order CE becomes similar to the 2$^{\text{nd}}$ CE curve when $m_2/m_1 \gtrsim4$, at least when $\sigma_{22}^{\text{T}}=5 \sigma_{11}^{\text{T}}$ and $\sigma_{12}^{\text{T}}$ has an intermediate value.

If we now compare both curves with the Green-Kubo results, we note that the 2$^{\text{nd}}$ order CE reproduces the qualitative behavior of the GK data better than the 1$^{\text{st}}$ order does, as the minimum of each full line is non-zero, while the dashed curves get to zero as $m_1=m_2$.  We also note that the relative discrepancy between 2$^{\text{nd}}$ order CE and GK is smaller for higher masses $m_1$ at fixed $m_2$, but also for higher $m_2$ at fixed $m_1$. In particular, for each choice of $m_1$ we see that the maximum relative difference occurs for $m_2\simeq m_1$, i.e. around the minimum of each curve. We may interpret this by noting that for $m_1\sim m_2$ only the contributions from the 2$^{\text{nd}}$ order onwards are non-zero, so the difference may be attributed to the missing contribution of the higher orders in the CE approximation.

\section{Conclusions and Outlook}
\label{sec:concl}

In this work we have investigated the bulk viscosity of a binary mixture within a relativistic kinetic theory approach. Starting from an expansion of the Boltzmann distribution function around equilibrium, we have derived an analytical expression for the bulk viscosity of a mixture $\zeta_{\text{mix}}^{(2)}$ at the 2$^\text{nd}$ order in the Chapman-Enskog expansion. In particular the main result of this work, i.e. Eq. \eqref{eq:zeta_2comp_2order}, significantly improves the well-known 1$^\text{st}$ order result, mainly because only at the 2$^\text{nd}$ order the effect of interactions among particles of the same species is included. Indeed, the 1$^\text{st}$ order Chapman-Enskog result depends only on the inter-species cross section $\sigma_{12}$. In addition to that, it vanishes if the difference of the two masses is zero. Physically speaking, this occurs because the viscosity of a mixture with equal masses, as well as the contribution coming from the intra-species cross sections ($\sigma_{11}$ and $\sigma_{22}$), are sub-leading in the Chapman-Enskog expansion. By evaluating the 2$^\text{nd}$ result in Eq. \eqref{eq:zeta_2comp_2order} we have been able to describe these two physical features, which in some instances can even be dominant. Moreover, this new formula reduces correctly to the $\zeta$ of a homogeneous fluid when we consider two indistinguishable components. Interestingly, we find that the first order CE approximation for the mixture reduces to $\zeta=0$ in the homogeneous limit, while the second order CE approximation for the mixture reduces to the first order homogeneous formula. This raises the question on whether the generic $n^{\text{th}}$ order CE for a mixture reduces to the $(n-1)^{\text{th}}$ order CE for the single-component gas, when one performs the homogeneous limit. To this day, there is no answer to such question.

Both the first and the second order Chapman-Enskog results have been compared against an independent numerical calculation using the Boltzmann transport equation, by which one can estimate $\zeta$ via the Green-Kubo formula. Using the GK data as a benchmark, we see that by going to higher orders in the CE expansion the agreement improves significantly, especially when the masses of the two components and/or their difference is small.

This work sheds light on the relative importance of all the different scales coming into play in the determination of the bulk viscosity of binary mixtures. Moreover, it provides an analytical framework which takes them all into account. The purpose of the paper is general, and such are its possible generalizations. In light of applications to the study of heavy-ion collisions, one may consider the inclusion of thermal-dependent masses, as done in \cite{Sambataro:2024mkr} and works cited therein. This would allow for a proper treatment of the non-perturbative effects within the Quark-Gluon Plasma. Moving instead to the study of neutron stars, here Pauli blocking effects are quite substantial \cite{Typel:2009sy}, therefore a future extension of this study to include these effects has to take into account a Boltzmann equation with the inclusion of the Uehling–Uhlenbeck terms in the collision integral \cite{Wang:2023gta}. Notice that, on the other hand, the collective evolution of the Quark-Gluon Plasma may be satisfactorily pictured as that of a Boltzmann-like system \cite{Heinz:2024jwu}.
%\corange{footnote{{We notice instead that a suitable study of the collective motion of the Quark-Gluon Plasma does not necessarily require the inclusion of quantum statistics, since its collective evolution can be successfully pictured as that of a classical system \cite{Heinz:2024jwu}. }}}
Finally, for a fully realistic study of both Quark-Gluon Plasma and neutron stars, it is necessary to take into account for the inclusion of flavor- and number-changing processes, as done e.g. in \cite{Ohanaka:2026hjx}. Indeed, these make up the largest contribution in these physical environments. We reserve these topics for future work.

\subsection*{Data availability}
The data that support the findings of this article are openly available in \cite{data_folder}.

\subsection*{Acknowledgments}
We thank A. Dash for useful discussions. G.P. and V.N. thank O. Vanoni for inspiration.

\appendix

\onecolumngrid

\section{Details of the CE calculations for \texorpdfstring{$\zeta$}{zeta}}
\label{sec:appendix_details_of_calculations}
In this Appendix we give more details on the explicit evaluation of the bulk viscosity from the Chapman-Enskog expansion. For the full theoretical treatment of the CE expansion starting from the Boltzmann equation check \cite{VANLEEUWEN197365,VANLEEUWEN1975233}. Here we limit ourselves to report the operationally useful formulae needed for this purpose.\\

The bulk viscosity $\zeta_{\text{mix}}^{(p)}$ for a binary mixture at the $p^{\text{th}}$ order in the CE approximation is given by
\begin{equation}
    \zeta_{\text{mix}}^{(p)}=nT\sideset{}{'}\sum_{s=-p-1}^{p+1} a_s^{(p)} \alpha_s.
    \label{eq:details_calculation_zeta_pthorder}
\end{equation}
The prime denotes that the $s=0$ term is omitted from the summation. In this expression there appear the terms $\alpha_s$, which are given by \cite{VANLEEUWEN197365}
\begin{equation}
\alpha_s=-\frac{1}{nT}\sum_{i=1}^2\theta_{si}
\sum_{j=1}^{|s|-1}\frac{z_i^j}{j!}L_{|s|-1-j}^{1/2+j}(-z_i)\,u_{\nu_1\nu_2\,\dots\, \nu_j}Q_i^{\nu_1\nu_2\,\dots\, \nu_j},
    \label{eq:details_calculation_alpha}
\end{equation}
where $L_{n}^\alpha (x)$ is the generalized Laguerre polynomial
\begin{equation}
    L_{n}^\alpha(x)\equiv \sum_{m=0}^n \binom{n+\alpha}{n-m}\frac{(-x)^m}{m!},
    \label{eq:details_calculation_Laguerre}
\end{equation}
and
\begin{equation}
\theta_{si}=\theta(s)\delta_{1i}+\theta(-s)\delta_{2i},
    \label{eq:details_calculation_thetasi}
\end{equation}
being $\theta(s)$ the Heaviside step function. Moreover
\begin{equation}
    Q_i^{\nu_1\nu_2\,\dots\, \nu_j}=\frac{1}{m_i^j}\int p_i^{\nu_1} \cdots p_i^{\nu_j}\,Q_i f_{i}^{(0)}\frac{\dd^3 p_i}{p_0^i},
    \label{eq:details_calculation_Qnu1nuj}
\end{equation}
\begin{equation}
Q_i=-\frac13 m_i^2+p_\alpha^i U^\alpha\left[(1-\gamma)m_i\hat{h}_i+\gamma T\right]+\left(\frac43 -\gamma \right) (p_\alpha^i U^\alpha)^2,
\label{eq:details_calculation_Q}
\end{equation}

\begin{equation}u^{\nu_1\nu_2\,\dots\, \nu_j}=U^{\nu_1}U^{\nu_2}\cdots U^{\nu_j},
\label{eq:details_calculation_unu1nun}
\end{equation}
being $U^{\mu}$ the hydrodynamic four velocity such that $U^{\mu}U_{\mu}=-1$, and $f_i^{(0)}$ the equilibrium distribution function of the fluid component $i$. To perform the integrations in \eqref{eq:details_calculation_Qnu1nuj} one should know the moments
\begin{equation}
F^{\nu_1\nu_2\,\dots\, \nu_j}=\int p^{\nu_1} \cdots p^{\nu_j} f^{(0)}\frac{\dd^3 p}{p_0},
\label{eq:details_calculation_momentsF}
\end{equation}
which are easily calculable and explicitly listed up to $j=6$ in \cite{VANLEEUWEN197365}. Finally, to express the results for $\alpha_s$ in the form presented in \eqref{eq:zeta_2ndorder_alpha2}, \eqref{eq:zeta_2ndorder_alpha3} and \eqref{eq:zeta_2ndorder_alpha-3}, the following identity is useful
\begin{equation}
\frac{\gamma_i}{\gamma_i-1}=z_i^2\left(1-\hat{h}_i^2+5z_i^{-1} \hat{h}_i\right),
    \label{eq:details_calculation_identity}
\end{equation}
which follows from Eq. \eqref{eq:twocompzeta_1storder_gamma_gamma1} applied for each component $i$ \cite{VANLEEUWEN197365}.\\

The coefficients $a_s^{(p)}$ in \eqref{eq:details_calculation_zeta_pthorder} cannot be derived directly, but they have to be found by inverting, at the $p^\text{th}$-order in the CE approximation, the following system of equations
\begin{equation}
    \sideset{}{'}\sum_{s=-p-1}^{p+1} a_{rs}a_s^{(p)}=n^{-1}\alpha_r.
    \label{eq:details_calculation_as(p)_coefficients}
\end{equation}
This would give back $a_s^{(p)}$ in terms of the coefficients $a_{rs}$, which encode the interactions among the particles in the fluid. They have the following form:
\begin{equation}
a_{rs}=a_{rs}'+a_{rs}''.
    \label{eq:details_calculation_ars}
\end{equation}
The single primed coefficients refer to interactions among particles within each species, and their expression is of the form:
\begin{equation}
    a_{rs}'=\sum_{i=1}^{2} x_i^2\,\theta_{rsi}\, \bar{a}_{|r|-1,|s|-1},
\end{equation}
where $\bar{a}_{m,n}$ are the single-component coefficients, explicitly listed in \cite{VANLEEUWEN197365} (for instance, $\bar{a}_{2,2}$ has the expression in Eq. \eqref{eq:zeta_1comp_parameters}), and the $\theta_{rsi}$ is
\begin{equation}
    \theta_{rsi}=\theta(r)\theta(s)\delta_{1i}+\theta(-r)\theta(-s)\delta_{2i}.
\end{equation}
On the other hand, the double primed coefficients $a_{rs}^{''}$ describe interaction between particles of species 1 with particles of species 2, and their expression is \cite{VANLEEUWEN1975233}
\begin{equation}
a_{rs}''=\frac{x_1 x_2 z_1 z_2}{2K_2(z_1)K_2(z_2)}\sum_{i,j=1}^{2}\theta_{rsij}\frac{1}{(|r|-1)!(|s|-1)!}[\partial_u^{|r|-1}\partial_{v}^{|s|-1} A_{ij}]_{u=v=0},
    \label{eq:details_calculation_a''_rs}
\end{equation}
where
\begin{equation}
\theta_{rsij}=\theta(r)\theta(s)\delta_{1i}\delta_{1j}+\theta(r)\theta(-s)\delta_{1i}\delta_{2j}+\theta(-r)\theta(s)\delta_{2i}\delta_{1j}+\theta(-r)\theta(-s)\delta_{2i}\delta_{2j},
\label{eq:details_calculation_thetarsij}
\end{equation}
\begin{align}
A_{ij}=&\left(\frac{U}{u}\right)^{3/2}\left(\frac{V}{v}\right)^{3/2}e^{z_i U+z_j V}\sum_{l=0}^{+\infty}\sum_{k=0}^{2l}\frac{(2l+1)!!}{k! (2k-l)!}\int\left[U(-1)^i z_i\sinh\psi_i\right]^k\left[V(-1)^j z_j\sinh\psi_j\right]^{2l-k}\nonumber\\
&\cdot\frac{K_{l+1}(W_{ij})}{W_{ij}^{l+1}}\left[f_{2l,0}(\Theta_{12})-f_{2l-k,k}(\Theta_{12})\right]\sigma_{12}(\Psi_{12},\Theta_{12})\sinh^3\Psi_{12}\dd \Psi_{12}\sin\Theta_{12}\dd\Theta_{12},
    \label{eq:details_calculation_Aij}
\end{align}
\begin{equation}
U=\frac{u}{1-u},~~~V=\frac{v}{1-v},~~~W_{ij}=\frac{P_{12}}{T}+U z_i \cosh\psi_i+V z_j \cosh\psi_j,
    \label{eq:details_calculation_U_V_Wij}
\end{equation}
\begin{equation}
f_{mn}(\Theta_{12})=\frac{4\pi \,m!\, n!\,[(m+n)/2]!}{(m+n+1)!}\sum_{k=0}^{\lfloor\min(m,n)/2\rfloor}\frac{(2\cos\Theta_{12})^{\min(m,n)-2k}}{k!\, (k+|m-n|/2)!\,[\min(m,n)-2k]!}.
    \label{eq:details_calculation_f_mntheta}
\end{equation}

Their definition implies that $a_{rs}'=a_{sr}'$ and $a_{rs}''=a_{sr}''$, therefore $a_{rs}=a_{sr}$. Moreover, from the conservation of energy and momentum it follows that
\begin{equation}
a_{1,r}=0,~~~~a_{2,r}'=0~~(r=\pm1,\pm2,\dots),~~~~a_{2,2}''=a_{-2,-2}''=-a_{2,-2}''.
    \label{eq:details_calculation_ars_identities}
\end{equation}
The coefficients $a_{rs}'$ are expressed in terms of omega integrals \eqref{eq:single_component_omegaintegrals}, similarly $a_{rs}''$ can be expressed as sums of generalized omega integrals of the form in Eq. \eqref{eq:omega_rtuv}. By doing so, one derives Eqs. \eqref{eq:zeta_2ndorder_a22} to \eqref{eq:zeta_2ndorder_a3-3} for the coefficients $a_{rs}$.

\section{Analytical calculations for the 2-component to 1-component reduction of \texorpdfstring{$\zeta$}{zeta}}
\label{sec:appendix_reduction2to1comp}

As argued in the main text, the $1^{\text{st}}$ order CE approximation does not capture the full physical behavior of the bulk viscosity of a mixture. In particular, this result does not reduce to the correct homogeneous formula \eqref{eq:zeta_1comp_1storder} when we consider all the masses and cross sections equal. On the other hand, we will show that the $2^\text{nd}$ order CE formula \eqref{eq:zeta_2comp_2order} for $\zeta$ reduces to the homogeneous $1^\text{st}$ order CE formula \eqref{eq:zeta_1comp_1storder} in this limit. \\

Let us assume that $m_1=m_2\equiv m$ and $\sigma_{11}=\sigma_{22}=\sigma_{12}\equiv\sigma$. Then, it is easy to see that equations \eqref{eq:zeta_2ndorder_alpha2} to \eqref{eq:zeta_2ndorder_a3-3} reduce to:

\begin{align}
    \alpha_2=&0,\label{eq:zeta_2ndorder_alpha2_reduced}\\
    \alpha_3=&\frac32x_1\left[z\hat{h}\left(\gamma-\frac53\right)+\gamma\right],\label{eq:zeta_2ndorder_alpha3_reduced}\\
    \alpha_{-3}=&\frac32x_2\left[z\hat{h}\left(\gamma-\frac53\right)+\gamma\right],\label{eq:zeta_2ndorder_alpha-3_reduced}\\
    a_{2,2}=&4x_1x_2\omega_{1200}^{(1)},\label{eq:zeta_2ndorder_a22_reduced}\\
    a_{2,3}=&x_1x_2\left[(10+4z)\,\omega_{1200}^{(1)}-4z\,\omega_{1210}^{(1)}\right],\label{eq:zeta_2ndorder_a23_reduced}\\
    a_{-2,-3}=&x_1 x_2\left[(10+4z)\,\omega_{1200}^{(1)}-4z\,\omega_{1201}^{(1)}\right]=a_{2,3},\label{eq:zeta_2ndorder_a-2-3_reduced}\\
    a_{3,3}=&2x_1^2\,\omega_{0}^{(2)}+x_1 x_2\left[\omega_{2300}^{(2)}+(5+2z)^2\omega_{1200}^{(1)}+4z^2\, \omega_{1220}^{(1)}+10z\,\omega_{1320}^{(1)}-4z(5+2z)\omega_{1210}^{(1)}\right]\nonumber\\
    \equiv&2x_1^2\,\omega_{0}^{(2)}+x_1 x_2\omega_{2300}^{(2)}+A,\label{eq:zeta_2ndorder_a33_reduced}\\
    a_{-3,-3}=&2x_2^2\,\omega_{0}^{(2)}+x_1 x_2\left[\omega_{2300}^{(2)}+(5+2z)^2\omega_{1200}^{(1)}+4z^2\, \omega_{1202}^{(1)}+10z\,\omega_{1302}^{(1)}-4z(5+2z)\omega_{1201}^{(1)}\right]\nonumber\\
    \equiv&2x_2^2\,\omega_{0}^{(2)}+x_1 x_2\omega_{2300}^{(2)}+A,\label{eq:zeta_2ndorder_a-3-3_reduced}\\
    a_{3,-3}=&x_1 x_2\left[\omega_{2300}^{(2)}-(5+2z)^2\omega_{1200}^{(1)}-4 z^2 \omega_{1211}^{(1)}-10z\,\omega_{1311}^{(1)}+4z(5+2z)\omega_{1201}^{(1)}\right]\nonumber\\
    \equiv&x_1x_2\omega_{2300}^{(2)}-A,\label{eq:zeta_2ndorder_a3-3_reduced}
\end{align}
where we noticed that when the masses are equal we have
\begin{equation}
\omega_{1220}^{(1)}=\omega_{1202}^{(1)}=\omega_{1211}^{(1)},~~~\omega_{1320}^{(1)}=\omega_{1302}^{(1)}=\omega_{1311}^{(1)},~~~
    \omega_{1210}^{(1)}=\omega_{1201}^{(1)},
    \label{eq:omega_integrals_reduction1}
\end{equation}
and we have defined
\begin{equation}
    A\equiv x_1x_2\left[(5+2z)^2\omega_{1200}^{(1)}+4z^2\, \omega_{1220}^{(1)}+10z\,\omega_{1320}^{(1)}-4z(5+2z)\omega_{1210}^{(1)}\right].
    \label{eq:A_definition}
\end{equation}
Under these assumptions, and by considering that in this limit $\omega_{2300}^{(2)}=\omega_0^{(2)}$ (check for instance Eq. (A5) in \cite{Parisi:2025gwq} and calculations therein), the denominator $\mathcal{D}$ in \eqref{eq:denominator_D} reduces to:
\begin{align}
    \mathcal{D}%\equiv& a_{2,2}(a_{3,-3}^2-a_{3,3}\,a_{-3,-3})+a_{2,3}^2\, a_{-3,-3}+a_{-2,-3}^2\,a_{3,3}+2a_{2,3}\,a_{3,-3}\,a_{-2,-3}=\nonumber\\
    =&a_{2,2}\left[A^2-2x_1x_2 A\, \omega_0^{(2)}+x_1^2x_2^2\left(\omega_0^{(2)}\right)^2-4x_1^2x_2^2 \left(\omega_0^{(2)}\right)^2-2x_1^3x_2 \left(\omega_0^{(2)}\right)^2\nonumber\right.\\
    &\left.-2x_1^2A\,\omega_0^{(2)}-2x_1x_2^3\left(\omega_0^{(2)}\right)^2-x_1^2 x_2^2\left(\omega_0^{(2)}\right)^2-x_1x_2 A\, \omega_0^{(2)}-2x_2^2 A\,\omega_0^{(2)}-x_1x_2 A\, \omega_0^{(2)}-A^2\right]\nonumber\\
    &+a_{2,3}^2\left[2x_2^2\omega_0^{(2)}+x_1x_2 \omega_0^{(2)}+A+2x_1^2 \omega_0^{(2)}+x_1 x_2 \omega_0^{(2)}+A-2A+2x_1x_2\omega_0^{(2)}\right]\nonumber\\
    =&a_{2,2}\left[-2A\, \omega_0^{(2)}-2x_1x_2\left(\omega_0^{(2)}\right)^2\nonumber\right]+a_{2,3}^22\omega_0^{(2)}\nonumber\\
    =&2\omega_0^{(2)}\left[-a_{2,2}\left(A+x_1x_2\omega_0^{(2)}\right)+a_{2,3}^2\right],
\end{align}
where we have exploited that $x_1+x_2=1$, by definition of number fraction. Substituting this result into \eqref{eq:zeta_2comp_2order} and exploiting once again Eqs. \eqref{eq:zeta_2ndorder_alpha2_reduced} to \eqref{eq:zeta_2ndorder_a3-3_reduced}, we get:

\begin{align}
    \zeta_{\text{mix}}^{(2)}\to&\frac{T}{\mathcal{D}}\left\{\frac32 \left[z\hat{h}\left(\gamma-\frac53 \right)+\gamma\right]\right\}^2\left[(a_{2,3}^2-a_{2,2}\,a_{3,3})x_2^2+2(a_{2,2}\,a_{3,-3}+a_{2,3}\,a_{-2,-3})x_1x_2+(a_{-2,-3}^2-a_{2,2}\,a_{-3,-3})x_1^2\right]\nonumber\\
    =&\frac{T}{\mathcal{D}}\left\{\frac32 \left[z\hat{h}\left(\gamma-\frac53 \right)+\gamma\right]\right\}^2\left\{a_{2,3}^2+a_{2,2}\left[-x_2^2\left(2x_1^2\omega_0^{(2)}+x_1x_2\omega_0^{(2)}+A\right)\right.\right.\nonumber\\
    &+\left.\left.2x_1x_2\left(x_1x_2\omega_0^{(2)}-A\right)-x_1^2\left(2x_2^2\omega_0^{(2)}+x_1x_2\omega_0^{(2)}+A\right)
    \right]\right\}\nonumber\\
    =&\frac{T}{\mathcal{D}}\left\{\frac32 \left[z\hat{h}\left(\gamma-\frac53 \right)+\gamma\right]\right\}^2\left[a_{2,3}^2-a_{2,2}A-x_1x_2a_{2,2} \omega_0^{(2)} \right]\nonumber\\
    =&\frac{T\bar{\alpha}_2^2}{2\omega_0^{(2)}}\frac{a_{2,3}^2-a_{2,2}A-x_1x_2a_{2,2} \omega_0^{(2)}}{a_{2,3}^2-a_{2,2}A-x_1x_2a_{2,2}\omega_0^{(2)}}\nonumber\\
    =&\frac{\bar{\alpha}_2^2}{\bar{a}_{22}}T.
\end{align}
As mentioned, we see that when all the masses and cross sections in the mixture are equal, we recover the 1-component 1$^{\text{st}}$ order CE bulk viscosity as in \eqref{eq:zeta_1comp_1storder}. Moreover, the result does not depend anymore on the relative concentrations of the two species $x_1,x_2$, as expected since in this limit the two species are indistinguishable.

\section{Details on the numerical solution of the Boltzmann Equation}
\label{sec:appendix_numerical}

As stated in the main text (Section \ref{sec:GK}) the evaluation of the bulk viscosity is carried out by means of a version of a relativistic Boltzmann transport code which has been optimized for this purpose. Here we focus on the main features of this optimization, and refer to \cite{Nugara:2023eku} for more details on the code in general.\\

The goal of the code is to solve the relativistic Boltzmann transport equation by mapping the evolution of the phase space distribution function $f(x,p)$. Such distribution is sampled by $N$ test particles, whose momenta are evolved in order to provide a numerical solution of the Boltzmann equation. In our formalism we have the freedom to arbitrarily choose the particle number $N$, which we tune in order to reach numerical convergence of the Green-Kubo results. Once $N$ is fixed, the volume of the system is constrained as $V=N/n$, where $n$ is the equilibrium number density at the given value of temperature $T$ and particle mass $m$. For an accurate representation of the distribution function in the $T=$ 0.1--0.8 GeV regime, which is the range explored in this work, we have found that $N\approx3\,000$ already represents a suitable value. 

The momenta are sampled from the J\"uttner distribution using rejection sampling, with a Gamma-distributed proposal (envelope) for the momentum magnitude.
Since the number of particles is finite, we need to take into account for a spurious non-zero total momentum. We cure this numerical artifact by Lorentz-boosting the system to the Landau frame, and comparing our numerical $\langle p^2 \rangle$ with the theoretical average, that is
\[
\langle p^2 \rangle =
\begin{cases}
12T^2 & m = 0, \\[6pt]
3mT\dfrac{K_3(m/T)}{K_2(m/T)} & m \neq 0.
\end{cases}
\]
Any discrepancy, then, is taken into account by scaling the momenta of each particle by an appropriate factor, so to reproduce the above theoretical estimate.

Concerning the space discretization, for the sake of this work we can consider a single cell having the size of the whole simulation volume $V$. Indeed, since we are interested only in the evolution of momentum coordinates, which describe the relaxation to equilibrium of the bulk viscous pressure fluctuations, for this purpose the coordinate space positions of the particles are not relevant, and so they are not being tracked in our implementation. We assume periodic boundary conditions, so that particles can never escape the boundaries of the volume $V$ as the system evolves for the entirety of simulation time. Notice that, within this implementation, all the possible pairs of particles should be taken into account for collisions at each time. Instead, in order to avoid quadratic complexity $\mathcal{O}(N^2)$, at each time step we perform $\mathcal{O}(N)$ collision trials at random. The original collision rate of full quadratic pairing is then restored by enhancing the collision probabilities (see \eqref{eq:collision_probability} below) by a suitable constant factor. Specifically, we sample only $6\,000$ candidate pairs per timestep for $T \leq 0.6$ GeV, and increase this number to $12\,000$ trials for $T > 0.6$ GeV. Indeed, at higher temperatures additional collision trials are required to control statistical errors, therefore increasing the computational cost due to more extensive Monte Carlo sampling.

Let us now discuss time discretization. One should keep in mind that, at each time step, in the stochastic implementation of the collision integral, a collision for any given pair of particles occurs with probability
\begin{equation}
\mathcal{P}= \frac{\sigma^{\text{T}} v_{\text{rel}}\, \Delta t}{V}. \label{eq:collision_probability}
\end{equation}
In the above equation $\sigma^{\text{T}}$ is the total interaction cross section and $v_{\text{rel}}$ is the relative speed of the two particles. Given this, in our simulations the time step size $\Delta t$ has been chosen such that the average number of collisions at each time step is roughly 60: we found $2\% $ of the total number of test particles $N$ to be a good enough value to capture the shape of the correlator curve with a resolution of $N_{T_{\text{max}}}=500$ points. With 60 collisions occurring at each timestep, after $N_{T_{\text{max}}}=500$ timesteps we expect all $N=3\,000$ particles to have their momenta significantly changed with respect to their initial values. This ensures a proper exponential decay of the correlator. In order to find such optimal timestep, the candidate values for $\Delta t$ are tested one by one by test-running the system with a fixed number of time steps, until the desired collision rate is achieved. This can be done quickly by starting with a very small value for $\Delta t$, e.g. 0.001 fm, and then repeatedly doubling it for the next test. After fixing the optimal time step size, the simulation is run for a maximum number $N_{\mathcal T_{\text{max}}}$ of iterations: we use $N_{\mathcal T_{\text{max}}} \approx 30\,000$. Thus, the total simulation time is $\mathcal T_{\text{max}} = N_{\mathcal{T}_{\text{max}}}\, \Delta t$. Indeed, we have found that this value of $N_{\mathcal T_{\text{max}}}$ is large enough to appropriately damp each correlator.

Finally, for each set of physical parameters, the Green-Kubo points for $\zeta$ have been obtained as an average over 20 numerical events. The evaluation of a single Green-Kubo point requires approximately 5 minutes on a Ryzen 5 3600 CPU using a pure \texttt{C} implementation. In contrast, a CUDA-accelerated implementation on GPU executed on an RTX 3070 reduces the evaluation time to approximately 30 seconds. The parallel CUDA version assigns one thread per particle, enabling substantial performance gains through fine-grained parallelism. To avoid race conditions in the parallel implementation for the collision procedure, each particle is assigned a collision partner via a self-inverse hash function $f(k) = ((N-1)k + S)\bmod N$ for $k = 0,1,\dots,N-1$, where $S$ is a random seed. This form of the hash function explicitly satisfies the property $f(k) = f^{-1}(k)$, which ensures that paired threads can mutually identify and synchronize with each other, as desired.\\

\begin{figure}[ht]
    \centering
    \includegraphics[width=0.5\linewidth]{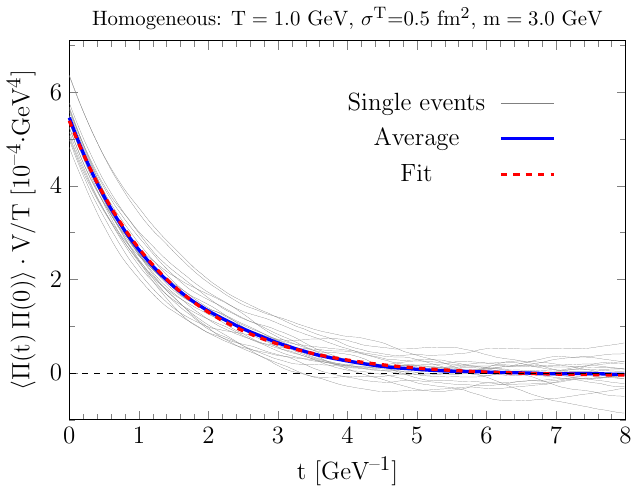}
    \caption{Time evolution of the (scaled) bulk pressure correlator for a single species interacting isotropically. Along with the single events reported as grey lines, we report their average (in blue) as well as an exponential fit (in red).}
    \label{fig:Bulk_correlator}
\end{figure}

In Figure \ref{fig:Bulk_correlator}, we present the single-event correlator calculations (thin gray lines), the ensemble-averaged correlator (thick blue line), and the curve obtained from an exponential fit over the average (dashed thick red line), for a single particle species with an isotropic cross section. Notice that the correlator is normalized by the factor $V/T$. With this normalization, the correlator becomes independent on the number of particles and depends only on the thermodynamic and interaction parameters, namely the temperature $T$, the total cross section $\sigma$ and the particle mass $m$.

\twocolumngrid

\bibliography{Bulk_viscosity_binary_mixture/biblio}

@article{Sambataro:2023tlv,
    author = "Sambataro, Maria Lucia and Minissale, Vincenzo and Plumari, Salvatore and Greco, Vincenzo",
    title = "{B meson production in Pb+Pb at 5.02 ATeV at LHC: Estimating the diffusion coefficient in the infinite mass limit}",
    eprint = "2304.02953",
    archivePrefix = "arXiv",
    primaryClass = "hep-ph",
    doi = "10.1016/j.physletb.2024.138480",
    journal = "Phys. Lett. B",
    volume = "849",
    pages = "138480",
    year = "2024"
}

@article{VANLEEUWEN1975233,
title = {On relativistic kinetic gas theory: XIV. Transport coefficients of a binary mixture. General expressions},
journal = {Physica A: Statistical Mechanics and its Applications},
volume = {79},
number = {2},
pages = {233-255},
year = {1975},
issn = {0378-4371},
doi = {https://doi.org/10.1016/0378-4371(75)90154-5},
url = {https://www.sciencedirect.com/science/article/pii/0378437175901545},
author = {W.A. {Van Leeuwen} and A.J. Kox and S.R. {de Groot}}
}

@article{VANLEEUWEN197365,
title = {On relativistic kinetic gas theory: IX. Transport coefficients for systems of particles with arbitrary interaction},
journal = {Physica},
volume = {63},
number = {1},
pages = {65-94},
year = {1973},
issn = {0031-8914},
doi = {https://doi.org/10.1016/0031-8914(73)90179-1},
url = {https://www.sciencedirect.com/science/article/pii/0031891473901791},
author = {W.A. {van Leeuwen} and P.H. Polak and S.R. {de Groot}}
}

@article{Parisi:2025gwq,
    author = "Parisi, Gabriele and Nugara, Vincenzo and Plumari, Salvatore and Greco, Vincenzo",
    title = "{Shear viscosity of a binary mixture for a relativistic fluid at high temperature}",
    eprint = "2510.20704",
    archivePrefix = "arXiv",
    primaryClass = "hep-ph",
    doi = "10.1103/qd2t-2sdp",
    journal = "Phys. Rev. D",
    volume = "113",
    number = "1",
    pages = "014001",
    year = "2026"
}

@article{Karsch:2007jc,
    author = "Karsch, Frithjof and Kharzeev, Dmitri and Tuchin, Kirill",
    title = "{Universal properties of bulk viscosity near the QCD phase transition}",
    eprint = "0711.0914",
    archivePrefix = "arXiv",
    primaryClass = "hep-ph",
    reportNumber = "BNL-NT-07-47, RBRC-703",
    doi = "10.1016/j.physletb.2008.01.080",
    journal = "Phys. Lett. B",
    volume = "663",
    pages = "217--221",
    year = "2008"
}

@article{Meyer:2007dy,
    author = "Meyer, Harvey B.",
    title = "{A Calculation of the bulk viscosity in SU(3) gluodynamics}",
    eprint = "0710.3717",
    archivePrefix = "arXiv",
    primaryClass = "hep-lat",
    reportNumber = "MIT-CTP-3881",
    doi = "10.1103/PhysRevLett.100.162001",
    journal = "Phys. Rev. Lett.",
    volume = "100",
    pages = "162001",
    year = "2008"
}

@article{Ryu:2015vwa,
    author = "Ryu, S. and Paquet, J. -F. and Shen, C. and Denicol, G. S. and Schenke, B. and Jeon, S. and Gale, C.",
    title = "{Importance of the Bulk Viscosity of QCD in Ultrarelativistic Heavy-Ion Collisions}",
    eprint = "1502.01675",
    archivePrefix = "arXiv",
    primaryClass = "nucl-th",
    doi = "10.1103/PhysRevLett.115.132301",
    journal = "Phys. Rev. Lett.",
    volume = "115",
    number = "13",
    pages = "132301",
    year = "2015"
}

@article{Alford:2011pi,
    author = "Alford, Mark G. and Mahmoodifar, Simin and Schwenzer, Kai",
    title = "{Viscous damping of r-modes: Large amplitude saturation}",
    eprint = "1103.3521",
    archivePrefix = "arXiv",
    primaryClass = "astro-ph.HE",
    doi = "10.1103/PhysRevD.85.044051",
    journal = "Phys. Rev. D",
    volume = "85",
    pages = "044051",
    year = "2012"
}

@article{Schenke:2020uqq,
    author = {Schenke, Bj{\"o}rn and Shen, Chun and Teaney, Derek},
    title = "{Transverse momentum fluctuations and their correlation with elliptic flow in nuclear collision}",
    eprint = "2004.00690",
    archivePrefix = "arXiv",
    primaryClass = "nucl-th",
    doi = "10.1103/PhysRevC.102.034905",
    journal = "Phys. Rev. C",
    volume = "102",
    number = "3",
    pages = "034905",
    year = "2020"
}

@article{Ohanaka:2026hjx,
    author = "Ohanaka, Okey and Lin, Zi-Wei",
    title = "{Shear viscosity of a massless quark-gluon gas in chemical equilibrium including all $2\leftrightarrow 2$ cross sections}",
    eprint = "2602.08155",
    archivePrefix = "arXiv",
    primaryClass = "hep-ph",
    month = "2",
    year = "2026"
}

@article{Jones:2001ya,
    author = "Jones, P. B.",
    title = "{Bulk viscosity of neutron star matter}",
    doi = "10.1103/PhysRevD.64.084003",
    journal = "Phys. Rev. D",
    volume = "64",
    pages = "084003",
    year = "2001"
}

@article{Weinberg:1971mx,
    author = "Weinberg, Steven",
    title = "{Entropy generation and the survival of protogalaxies in an expanding universe}",
    doi = "10.1086/151073",
    journal = "Astrophys. J.",
    volume = "168",
    pages = "175",
    year = "1971"
}

@article{Zimdahl:1996ka,
    author = "Zimdahl, Winfried",
    title = "{Bulk viscous cosmology}",
    eprint = "astro-ph/9601189",
    archivePrefix = "arXiv",
    doi = "10.1103/PhysRevD.53.5483",
    journal = "Phys. Rev. D",
    volume = "53",
    pages = "5483--5493",
    year = "1996"
}

@article{Wiranata:2009cz,
    author = "Wiranata, A. and Prakash, M.",
    editor = "Stankus, Paul and Silvermyr, David and Sorensen, Soren and Greene, Victoria",
    title = "{Bulk Viscosity of Interacting Hadrons}",
    eprint = "0906.5592",
    archivePrefix = "arXiv",
    primaryClass = "nucl-th",
    doi = "10.1016/j.nuclphysa.2009.09.023",
    journal = "Nucl. Phys. A",
    volume = "830",
    pages = "219C--222C",
    year = "2009"
}

@article{Dash:2019zwq,
    author = "Dash, Ashutosh and Samanta, Subhasis and Mohanty, Bedangadas",
    title = "{Transport coefficients for multicomponent gas of hadrons using Chapman-Enskog method}",
    eprint = "1905.07130",
    archivePrefix = "arXiv",
    primaryClass = "nucl-th",
    doi = "10.1103/PhysRevD.100.014025",
    journal = "Phys. Rev. D",
    volume = "100",
    number = "1",
    pages = "014025",
    year = "2019"
}

@book{DeGroot:1980dk,
    author = "De Groot, S. R. and Van Leeuwen, W. A. and Van Weert, C. G.",
    title = "{Relativistic } {Kinetic } {Theory. } {Principles}\ {and} {Applications}",
    year = "1980",
    publisher     = "North-Holland Publishing Company",
    address       = "Netherlands",
}

@article{Chapman_book_1974,
author = {Jehring, L. and Chapman, S. and Cowling, T. G.},
title = "The Mathematical Theory of Non-Uniform Gases. 3rd edition. Cambridge etc., Cambridge University Press 1990. XXIV, 422 pp., £ 19.50 P/b. ISBN 0-521-40844-X",
journal = "ZAMM - Journal of Applied Mathematics and Mechanics/Zeitschrift für Angewandte Mathematik und Mechanik",
volume = {72},
number = {11},
pages = {610-610},
doi = {https://doi.org/10.1002/zamm.19920721124},
url = "https://onlinelibrary.wiley.com/doi/abs/10.1002/zamm.19920721124",
year = {1992}
}

@article{Israel_1963,
author = {Israel,Werner },
title = {Relativistic Kinetic Theory of a Simple Gas},
journal = {Journal of Mathematical Physics},
volume = {4},
number = {9},
pages = {1163-1181},
year = {1963},
doi = {10.1063/1.1704047},

URL = { 
        https://doi.org/10.1063/1.1704047
    
},
eprint = { 
        https://doi.org/10.1063/1.1704047
    
}
}

@article{Wiranata:2012br,
    author = "Wiranata, Anton and Prakash, Madappa",
    title = "{Shear Viscosities from the Chapman-Enskog and the Relaxation Time Approaches}",
    eprint = "1203.0281",
    archivePrefix = "arXiv",
    primaryClass = "nucl-th",
    doi = "10.1103/PhysRevC.85.054908",
    journal = "Phys. Rev. C",
    volume = "85",
    pages = "054908",
    year = "2012"
}

@article{Prakash:1993bt,
    author = "Prakash, Madappa and Prakash, Manju and Venugopalan, R. and Welke, G.",
    title = "{Nonequilibrium properties of hadronic mixtures}",
    doi = "10.1016/0370-1573(93)90092-R",
    journal = "Phys. Rept.",
    volume = "227",
    pages = "321--366",
    year = "1993"
}

@book{Denicol:2021clh,
    author = "Denicol, Gabriel S. and Rischke, Dirk H.",
    title = "{Microscopic Foundations of Relativistic Fluid Dynamics}",
    doi = "10.1007/978-3-030-82077-0",
    isbn = "978-3-030-82075-6, 978-3-030-82077-0",
    year = "2021"
}

@article{Plumari:2012ep,
    author = "Plumari, S. and Puglisi, A. and Scardina, F. and Greco, V.",
    title = "{Shear Viscosity of a strongly interacting system: Green-Kubo vs. Chapman-Enskog and Relaxation Time Approximation}",
    eprint = "1208.0481",
    archivePrefix = "arXiv",
    primaryClass = "nucl-th",
    doi = "10.1103/PhysRevC.86.054902",
    journal = "Phys. Rev. C",
    volume = "86",
    pages = "054902",
    year = "2012"
}

@article{Green:1954ubq,
    author = "Green, Melville S.",
    title = "{Markoff Random Processes and the Statistical Mechanics of Time-Dependent Phenomena. II. Irreversible Processes in Fluids}",
    doi = "10.1063/1.1740082",
    journal = "J. Chem. Phys.",
    volume = "22",
    number = "3",
    pages = "398",
    year = "1954"
}

@article{Kubo_1957,
author = {Kubo ,Ryogo and Yokota ,Mario and Nakajima ,Sadao},
title = {Statistical-Mechanical Theory of Irreversible Processes. II. Response to Thermal Disturbance},
journal = {Journal of the Physical Society of Japan},
volume = {12},
number = {11},
pages = {1203-1211},
year = {1957},
doi = {10.1143/JPSJ.12.1203},
eprint = { 
        https://doi.org/10.1143/JPSJ.12.1203
    
}

}

@article{Scardina:2017ipo,
    author = "Scardina, Francesco and Das, Santosh K. and Minissale, Vincenzo and Plumari, Salvatore and Greco, Vincenzo",
    title = "{Estimating the charm quark diffusion coefficient and thermalization time from D meson spectra at energies available at the BNL Relativistic Heavy Ion Collider and the CERN Large Hadron Collider}",
    eprint = "1707.05452",
    archivePrefix = "arXiv",
    primaryClass = "nucl-th",
    doi = "10.1103/PhysRevC.96.044905",
    journal = "Phys. Rev. C",
    volume = "96",
    number = "4",
    pages = "044905",
    year = "2017"
}

@article{Wiranata:2013oaa,
    author = "Wiranata, Anton and Koch, Volker and Prakash, Madappa and Wang, Xin Nian",
    title = "{Shear viscosity of hadrons with K-matrix cross sections}",
    eprint = "1307.4681",
    archivePrefix = "arXiv",
    primaryClass = "hep-ph",
    doi = "10.1103/PhysRevC.88.044917",
    journal = "Phys. Rev. C",
    volume = "88",
    number = "4",
    pages = "044917",
    year = "2013"
}

@book{zubarev1996statistical,
  author={Zubarev, D.N. and Morozov, V.G. and R{\"o}pke, G.},
title={Relaxation and hydrodynamic processes},
  isbn={9783055017094},
  lccn={97206642},
  series={Statistical Mechanics of Nonequilibrium Processes},
  year={1996},
  publisher={Akademie Verlag}
}

@article{Xu:2004mz,
    author = "Xu, Zhe and Greiner, Carsten",
    title = "{Thermalization of gluons in ultrarelativistic heavy ion collisions by including three-body interactions in a parton cascade}",
    eprint = "hep-ph/0406278",
    archivePrefix = "arXiv",
    doi = "10.1103/PhysRevC.71.064901",
    journal = "Phys. Rev. C",
    volume = "71",
    pages = "064901",
    year = "2005"
}

@article{Wang:2023gta,
    author = "Wang, Rui and Ma, Yu-Gang and Chen, Lie-Wen and Ko, Che Ming and Sun, Kai-Jia and Zhang, Zhen",
    title = "{Kinetic approach of light-nuclei production in intermediate-energy heavy-ion collisions}",
    eprint = "2305.02988",
    archivePrefix = "arXiv",
    primaryClass = "nucl-th",
    doi = "10.1103/PhysRevC.108.L031601",
    journal = "Phys. Rev. C",
    volume = "108",
    number = "3",
    pages = "L031601",
    year = "2023"
}

@article{Typel:2009sy,
    author = "Typel, S. and Ropke, G. and Klahn, T. and Blaschke, D. and Wolter, H. H.",
    title = "{Composition and thermodynamics of nuclear matter with light clusters}",
    eprint = "0908.2344",
    archivePrefix = "arXiv",
    primaryClass = "nucl-th",
    doi = "10.1103/PhysRevC.81.015803",
    journal = "Phys. Rev. C",
    volume = "81",
    pages = "015803",
    year = "2010"
}

@article{Nugara:2023eku,
    author = "Nugara, Vincenzo and Plumari, Salvatore and Oliva, Lucia and Greco, Vincenzo",
    title = "{Far-from-equilibrium attractors with full relativistic Boltzmann approach in boost-invariant and non-boost-invariant systems}",
    eprint = "2311.11921",
    archivePrefix = "arXiv",
    primaryClass = "hep-ph",
    doi = "10.1140/epjc/s10052-024-13227-1",
    journal = "Eur. Phys. J. C",
    volume = "84",
    number = "8",
    pages = "861",
    year = "2024"
}

@article{Nugara:2024net,
    author = "Nugara, Vincenzo and Greco, Vincenzo and Plumari, Salvatore",
    title = "{Far-from-equilibrium attractors with Full Relativistic Boltzmann approach in 3+1D: moments of distribution function and~anisotropic flows $v_n$}",
    eprint = "2409.12123",
    archivePrefix = "arXiv",
    primaryClass = "hep-ph",
    doi = "10.1140/epjc/s10052-025-14029-9",
    journal = "Eur. Phys. J. C",
    volume = "85",
    number = "3",
    pages = "311",
    year = "2025"
}

@article{Sambataro:2024mkr,
    author = "Sambataro, Maria Lucia and Greco, Vincenzo and Parisi, Gabriele and Plumari, Salvatore",
    title = "{Quasi particle model vs lattice QCD thermodynamics: extension to $N_f=2+1+1$ flavors and momentum dependent quark masses}",
    eprint = "2404.17459",
    archivePrefix = "arXiv",
    primaryClass = "hep-ph",
    doi = "10.1140/epjc/s10052-024-13276-6",
    journal = "Eur. Phys. J. C",
    volume = "84",
    number = "9",
    pages = "881",
    year = "2024"
}

@article{Hernandez:2025zxw,
    author = {Hernandez, Jos{\'e} Luis and Manuel, Cristina and S{\"a}ppi, Saga and Tolos, Laura},
    title = "{Burgers equation for the bulk viscous pressure of quark matter}",
    eprint = "2507.00794",
    archivePrefix = "arXiv",
    primaryClass = "hep-ph",
    doi = "10.1103/3v42-8dpw",
    journal = "Phys. Rev. D",
    volume = "113",
    number = "1",
    pages = "014032",
    year = "2026"
}

@article{Sappi:2025tcx,
    author = {S{\"a}ppi, Saga},
    title = "{Bulk viscosity and $n$-component fluids}",
    eprint = "2507.12365",
    archivePrefix = "arXiv",
    primaryClass = "hep-ph",
    month = "7",
    year = "2025"
}

@article{Gavassino:2023eoz,
    author = "Gavassino, Lorenzo",
    title = "{Relativistic bulk viscous fluids of Burgers type and their presence in neutron stars}",
    eprint = "2304.05455",
    archivePrefix = "arXiv",
    primaryClass = "nucl-th",
    doi = "10.1088/1361-6382/ace587",
    journal = "Class. Quant. Grav.",
    volume = "40",
    number = "16",
    pages = "165008",
    year = "2023"
}

@misc{data_folder,
    note = {Parisi, G. (2026). Bulk viscosity of a binary mixture: the role of the intra-species interaction [Data set]. Zenodo. \url{https://doi.org/10.5281/zenodo.20430253}}
}

@article{Oliva:2020doe,
    author = "Oliva, Lucia and Plumari, Salvatore and Greco, Vincenzo",
    title = "{Directed flow of D mesons at RHIC and LHC: non-perturbative dynamics, longitudinal bulk matter asymmetry and electromagnetic fields}",
    eprint = "2009.11066",
    archivePrefix = "arXiv",
    primaryClass = "hep-ph",
    doi = "10.1007/JHEP05(2021)034",
    journal = "JHEP",
    volume = "05",
    pages = "034",
    year = "2021"
}

@article{Sambataro:2020pge,
    author = "Sambataro, Maria Lucia and Plumari, Salvatore and Greco, Vincenzo",
    title = "{Impact of off-shell dynamics on the transport properties and the dynamical evolution of Charm Quarks at RHIC and LHC temperatures}",
    eprint = "2005.14470",
    archivePrefix = "arXiv",
    primaryClass = "hep-ph",
    doi = "10.1140/epjc/s10052-020-08704-2",
    journal = "Eur. Phys. J. C",
    volume = "80",
    number = "12",
    pages = "1140",
    year = "2020"
}

@article{Plumari:2019hzp,
    author = "Plumari, Salvatore and Coci, Gabriele and Minissale, Vincenzo and Das, Santosh K. and Sun, Yifeng and Greco, Vincenzo",
    title = "{Heavy - light flavor correlations of anisotropic flows at LHC energies within event-by-event transport approach}",
    eprint = "1912.09350",
    archivePrefix = "arXiv",
    primaryClass = "hep-ph",
    doi = "10.1016/j.physletb.2020.135460",
    journal = "Phys. Lett. B",
    volume = "805",
    pages = "135460",
    year = "2020"
}

@article{Sun:2019gxg,
    author = "Sun, Yifeng and Plumari, Salvatore and Greco, Vincenzo",
    title = "{Study of collective anisotropies $v_2$ and $v_3$ and their fluctuations in $pA$ collisions at LHC within a relativistic transport approach}",
    eprint = "1907.11287",
    archivePrefix = "arXiv",
    primaryClass = "nucl-th",
    doi = "10.1140/epjc/s10052-019-7577-7",
    journal = "Eur. Phys. J. C",
    volume = "80",
    number = "1",
    pages = "16",
    year = "2020"
}

@article{Plumari:2015cfa,
    author = "Plumari, Salvatore and Guardo, Giovanni Luca and Scardina, Francesco and Greco, Vincenzo",
    title = "{Initial state fluctuations from mid-peripheral to ultra-central collisions in a event-by-event transport approach}",
    eprint = "1507.05540",
    archivePrefix = "arXiv",
    primaryClass = "hep-ph",
    doi = "10.1103/PhysRevC.92.054902",
    journal = "Phys. Rev. C",
    volume = "92",
    number = "5",
    pages = "054902",
    year = "2015"
}

@article{Puglisi:2014sha,
    author = "Puglisi, A. and Plumari, S. and Greco, V.",
    title = "{Electric Conductivity from the solution of the Relativistic Boltzmann Equation}",
    eprint = "1408.7043",
    archivePrefix = "arXiv",
    primaryClass = "hep-ph",
    doi = "10.1103/PhysRevD.90.114009",
    journal = "Phys. Rev. D",
    volume = "90",
    pages = "114009",
    year = "2014"
}

@article{Scardina:2012mik,
    author = "Scardina, F. and Colonna, M. and Plumari, S. and Greco, V.",
    title = "{Quark-to-gluon composition of the quark-gluon plasma in relativistic heavy-ion collisions}",
    eprint = "1202.2262",
    archivePrefix = "arXiv",
    primaryClass = "nucl-th",
    doi = "10.1016/j.physletb.2013.06.034",
    journal = "Phys. Lett. B",
    volume = "724",
    pages = "296--300",
    year = "2013"
}

@article{Ferini:2008he,
    author = "Ferini, G. and Colonna, M. and Di Toro, M. and Greco, V.",
    title = "{Scalings of Elliptic Flow for a Fluid at Finite Shear Viscosity}",
    eprint = "0805.4814",
    archivePrefix = "arXiv",
    primaryClass = "nucl-th",
    doi = "10.1016/j.physletb.2008.10.062",
    journal = "Phys. Lett. B",
    volume = "670",
    pages = "325--329",
    year = "2009"
}

@article{Rose:2020lfc,
    author = "Rose, J. -B. and Torres-Rincon, J. M. and Elfner, H.",
    title = "{Inclusive and effective bulk viscosities in the hadron gas}",
    eprint = "2005.03647",
    archivePrefix = "arXiv",
    primaryClass = "hep-ph",
    doi = "10.1088/1361-6471/abbc5c",
    journal = "J. Phys. G",
    volume = "48",
    number = "1",
    pages = "015005",
    year = "2020"
}

@inproceedings{Heinz:2024jwu,
    author = {Heinz, Ulrich and Schenke, Bj{\"o}rn},
    title = "{Hydrodynamic Description of the Quark-Gluon Plasma}",
    eprint = "2412.19393",
    archivePrefix = "arXiv",
    primaryClass = "nucl-th",
    month = "12",
    year = "2024"
}

\end{document}